\documentclass[amsmath,amssymb,ams, preprint, superscriptaddress,nofootinbib]{revtex4-2} %,twocolumn
\usepackage[a4paper,textwidth=17.9cm,textheight=26cm]{geometry}
\usepackage[english]{babel}
\usepackage{ulem}
\usepackage{siunitx} 

\addto\captionsenglish{}
\addto\captionsenglish{}

\usepackage{upgreek}
\usepackage{graphicx}

\renewcommand{\baselinestretch}{1.5} 

\usepackage[T1]{fontenc}
\usepackage{lmodern}
\usepackage{helvet}
\usepackage{lineno}
\usepackage{xr}
\makeatletter
\newcommand*{\addFileDependency}[1]{
  \typeout{(#1)}
  \@addtofilelist{#1}
  \IfFileExists{#1}{}{\typeout{No file #1.}}
}
\usepackage{pdfpages} 
\AtBeginDocument{\let\LS@rot\@undefined}
\makeatother

\begin{document}
\title{Photonics-integrated terahertz transmission lines}

\author{Yazan Lampert}
 \altaffiliation{Contributed equally to this work}
\affiliation{Hybrid Photonics Laboratory, École Polytechnique Fédérale de Lausanne (EPFL),  CH-1015, Switzerland}
 \affiliation{Center for Quantum Science and Engineering (EPFL), CH-1015, Switzerland}
 
\author{Amirhassan Shams-Ansari}
 \altaffiliation{Contributed equally to this work}
 \affiliation{Harvard John A. Paulson School of Engineering and Applied Sciences, Harvard University, Cambridge, MA, USA}
  \affiliation{DRS Daylight Solutions, 16465 Via Esprillo, CA, USA}

  \author{Aleksei Gaier}
\affiliation{Hybrid Photonics Laboratory, École Polytechnique Fédérale de Lausanne (EPFL),  CH-1015, Switzerland}
 \affiliation{Center for Quantum Science and Engineering (EPFL), CH-1015, Switzerland}

\author{Alessandro Tomasino}
\affiliation{Hybrid Photonics Laboratory, École Polytechnique Fédérale de Lausanne (EPFL),  CH-1015, Switzerland}
 \affiliation{Center for Quantum Science and Engineering (EPFL), CH-1015, Switzerland}

 \author{Shima Rajabali}
\affiliation{Hybrid Photonics Laboratory, École Polytechnique Fédérale de Lausanne (EPFL),  CH-1015, Switzerland}
 \affiliation{Center for Quantum Science and Engineering (EPFL), CH-1015, Switzerland}
\affiliation{Harvard John A. Paulson School of Engineering and Applied Sciences, Harvard University, Cambridge, MA, USA}

 \author{Leticia Magalhaes}
\affiliation{Harvard John A. Paulson School of Engineering and Applied Sciences, Harvard University, Cambridge, MA, USA}

\author{Marko Lončar}
 \affiliation{Harvard John A. Paulson School of Engineering and Applied Sciences, Harvard University, Cambridge, MA, USA}
 
\author{Ileana-Cristina Benea-Chelmus}
\affiliation{Hybrid Photonics Laboratory, École Polytechnique Fédérale de Lausanne (EPFL),  CH-1015, Switzerland}
 \affiliation{Center for Quantum Science and Engineering (EPFL), CH-1015, Switzerland}
 
\date{\today}

%\linenumbers

\begin{abstract}
Modern communication and sensing technologies rely on connecting the optical domain with the microwave domain. Terahertz  technologies, spanning increased frequencies from 100 GHz to 10 THz, are critical for providing larger bandwidths and faster switching capabilities. Despite progress in high-frequency electronic sources and detectors, these technologies lack a direct link to the optical domain, and face challenges with increasing frequencies ($>$ 1~THz). Nonlinear processes, such as optical rectification, offer potential solutions for terahertz generation but are currently limited to discrete components based on bulk nonlinear crystals, missing out miniaturisation opportunities from integration of optical circuits with terahertz ones. We address this challenge by integrating phase-matched terahertz transmission lines with photonic circuits in a single hybrid architecture on the thin-film lithium niobate~(TFLN) platform. This design demonstrates phase-matched, broadband terahertz emission spanning four octaves (200 GHz to 3.5~THz) through broadband down-conversion of optical signals at telecommunication wavelengths. The micron-sized transmission lines provide terahertz field confinement with minimal radiative loss, enabling compact THz cavities embedded in integrated photonic circuits. This integration is crucial for leveraging photonics' advantages in low-noise, low-loss, and high-speed operations for terahertz generation. Our platform can readily be integrated with other mature photonics components, such as electro-optic modulators, frequency comb sources, or femtosecond sources to pave the way for compact, power-efficient, and frequency-agile broadband sources with applications in telecommunications, spectroscopy, quantum electrodynamics and optical computing.

\end{abstract}

\maketitle

\section*{Introduction}\label{sec1}
Terahertz (THz) technologies, with operating frequencies between 0.1-10 THz, hold transformative potential across a broad range of applications, including high-speed communication (6G technology)\cite{dan2019terahertz,chen2021terahertz, jiang2024terahertz}, non-destructive imaging~\cite{Jansen:10, Mittleman:18,shrekenhamer2013terahertz,usami2002development} and sensitive spectroscopy~\cite{charron2013chemical,browne2020prediction}. However, despite tremendous progress made, high-frequency electronic sources and detectors~\cite{sengupta_terahertz_2018}, such as multiplier chains~\cite{7563842} or plasma discharge~\cite{samizadeh2023electronic} lose efficiency with increasing frequencies, accentuating the need for efficient detectors and emitters. Given these tremendous challenges, one attractive approach is to detect and generate these high frequencies  through non-electronic techniques such as nonlinear optical mixing ~\cite{integrated_THz,sun_integrated_2024}. 

The integration of terahertz systems within integrated photonic structures is particularly appealing because they offer smaller size, weight, and power (SWaP). By combining these benefits with other engineering knobs such as geometric dispersion engineering and phase matching, nonlinear frequency mixing has the potential to provide a flexible and power-efficient method to generate and detect terahertz radiation \cite{boyd2008nonlinear}. Furthermore, the large analog bandwidth and the wide tuning range of photonics could provide effective solutions to modulate terahertz radiation at low drive voltages and with high linearity, faithfully following the optical drive signals. 

\begin{figure}[h!]
    \centering
    \includegraphics[width=17cm]{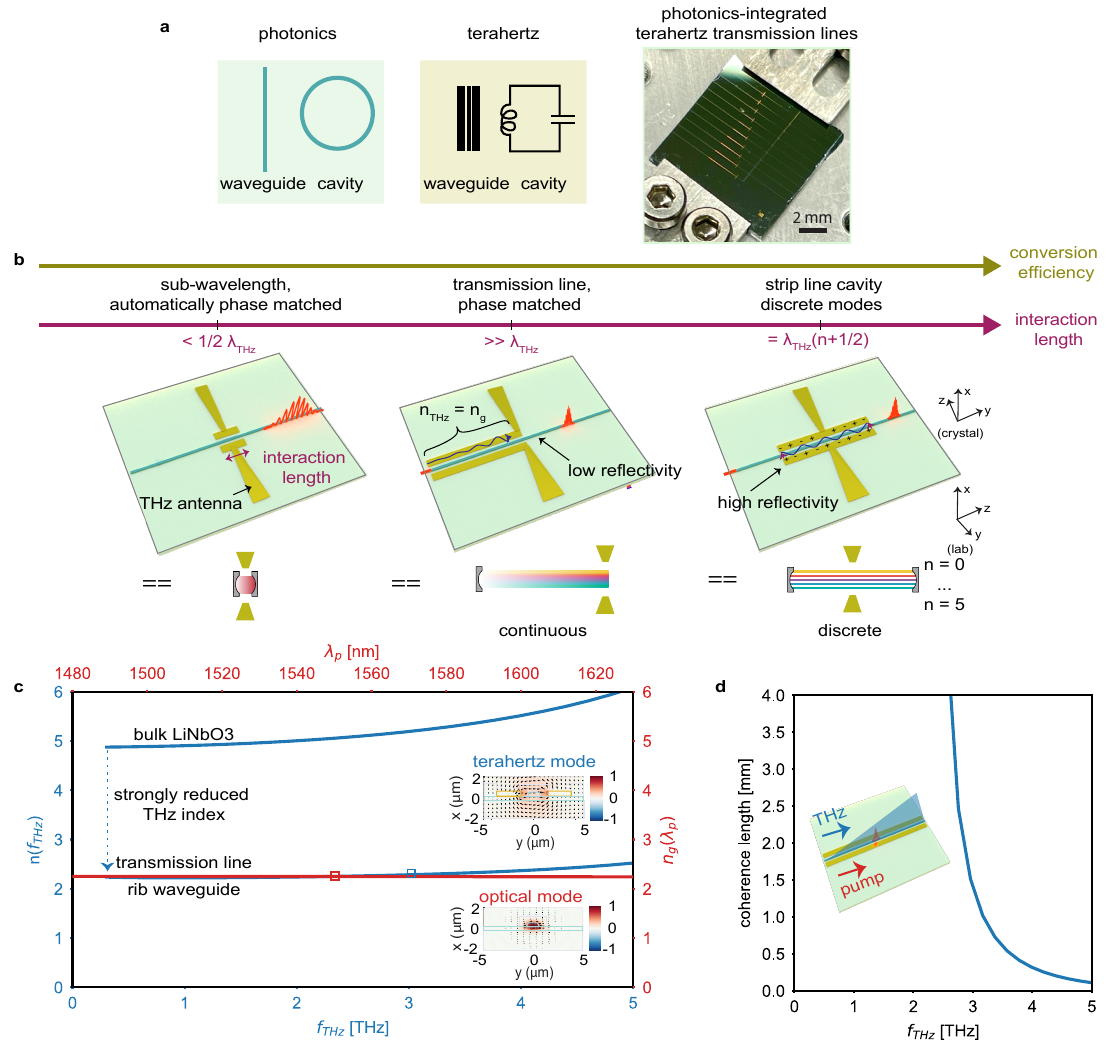}
    \renewcommand{\baselinestretch}{1} 
    \caption{\textbf{Integration of terahertz and photonics on a single chip.} \textbf{a} Waveguides and cavities are ubiquitous in photonics. In contrast, on-chip terahertz waveguides and cavities are little explored, limiting efficient exploration of nonlinear effects for terahertz generation or detection. Photograph of our proposed chip that integrates photonics with terahertz transmission lines. \textbf{b} Efficient generation of terahertz waves by optical rectification requires matching the terahertz effective index $\mathrm{n_{TL}}$ with the group index $\mathrm{n_{g}}$ of a telecom pulse propagating inside a rib waveguide. In the absence of phase matching, interaction lengths need to be limited to sub-terahertz-wavelengths ($\mathrm{l_{int} <\frac{1}{2}\lambda_{THz}}$, left panel) restricting the efficiency~\cite{herter2023}. In contrast, phase-matching enables an extended interaction length ($\mathrm{l_{int} >>\frac{1}{2}\lambda_{THz}}$, center panel), and a stronger build-up of the terahertz with a power scaling of $\mathrm{P_{THz} = \eta \sim l_{int}^2}$.  Terahertz transmission lines can achieve this, since their geometry can be chosen as to match the terahertz effective refractive index with the group index of light. In addition, an antenna can be patterned at the end of the transmission line to facilitate outcoupling into free space. Alternatively, the terahertz transmission line can be terminated with open ends on both sides to realise strip line cavities that support a discrete set of modes (on-resonance $\mathrm{l_{int} =(n+\frac{1}{2})\lambda_{THz}}$, right panel). \textbf{c} The effective refractive index of the fundamental mode at terahertz frequencies of the transmission line is matched to the group refractive index of the rib waveguide at telecom frequencies, while preserving good spatial overlap (inset). \textbf{d} Phase matching enables coherence lengths larger than a few millimeters in a co-linear, co-propagating configuration.}
    \renewcommand{\baselinestretch}{1.5} 
    \label{fig1}   
\end{figure}

Two key ingredients of any electromagnetic system are waveguides, used to route the signals, and cavities, needed to store and amplify the fields. These have been successfully implemented in integrated photonics ~\cite{nallappan2021terahertz}. However, developing effective methods to route and enhance terahertz waves in parallel with integrated photonic systems presents significant challenges~(Fig.~\ref{fig1}~a). These arise primarily from the wavelength mismatch between optics and terahertz waves, as well as the significant material absorption and dispersion at terahertz frequencies \cite{THz_science}.
Even leading nonlinear optical platforms with high $\chi^{(2)}$ nonlinearity and robust handling of high powers, such as Lithium Niobate (LN), exhibit strong dispersion and significant absorption at high terahertz frequencies. For instance, at 1~THz, LN's refractive index reaches $\mathrm{n(\omega_{THz}) = 5}$ with a dispersion parameter as high as $\mathrm{\beta_2 = 0.5\ ps^2/mm}$ and an absorption coefficient of $\mathrm{\alpha = 10~cm^{-1}}$~\cite{Wu:15}. This is in stark contrast to optical frequencies ($\omega_\text{p}$), where the refractive index of LN is significantly lower ($\mathrm{n(\omega_\text{p}) = 2.1}$), and remains relatively flat across these frequencies~\cite{Zelmon:97}. 

In bulk systems, these contrasting material properties have complicated the generation of broadband terahertz through processes such as optical rectification. Complex techniques such as non-collinear phase matching, terahertz generation at the Cherenkov angle~\cite{theuer_efficient_2006,l2007generation}, or tilted pulse front technique~\cite{guiramandNearoptimalIntensePowerful2022,10.1063/1.1617371,fulop2010design}  result in a non-Gaussian terahertz beam shape. On the other hand, collinear phase matching is possible either in ultra-thin crystals~\cite{carletti_nonlinear_2023}, offering limited output terahertz power, or in periodically poled lithium niobate, limiting the bandwidth for efficient terahertz-optical conversion \cite{lee2000generation,jolly2019spectral,lee2001tunable}. 
Some of the aforementioned problems have been addressed by placing crystals inside a laser cavity~\cite{wang2023high} or patterning of waveguides for the optical or terahertz signals, either experimentally~\cite{dastrup2022enhancement} or theoretically \cite{yang2021efficient}, including on thin-film lithium niobate (TFLN) platform \cite{zhu2021integrated}. In TFLN, the integration of LN rib waveguides with various structures of terahertz bowtie antennas have led to on-chip control over the emission frequency, amplitude, and time-domain waveform of the terahertz field \cite{herter2023}. On the downside, these configurations were limited by short interaction lengths—less than half a terahertz wavelength— which severely restricted the power and frequency range of the generated terahertz radiation (Fig.~\ref{fig1}b, left panel)\cite{herter2023}. 

Building on these efforts, we introduce a hybrid architecture for phase-matched integrated terahertz transmission lines on TFLN, enabling broadband terahertz generation (photograph in Fig.~\ref{fig1}~a and Fig.~\ref{fig1}~b). Our device consists of a rib waveguide inside a terahertz stripline to generate, couple, and guide terahertz fields effectively (Fig.~\ref{fig1}~b, center panel). This confines the terahertz to a strongly sub-wavelength cross-section ($\mathrm{\frac{S_{eff}}{\lambda^2} = 10^{-5}}$), thereby mitigating the associated dispersion and absorption losses. Similar to phase matching between radio-frequency fields and optical modes, we demonstrate the design's ability to achieve phase matching across an extensive frequency range from up to 3.5 THz~(Fig.~\ref{fig1}~c) with a coherence length with values of few  millimeters~(Fig.~\ref{fig1}~d) and obtain spatial overlap of the terahertz and optical modes (insets Fig.~\ref{fig1}~c). Altogether, this enables the generation of a broad band of terahertz frequencies spanning from 200~GHz up to 3.5~THz through optical rectification, and a field amplitude higher than current state-of-the-art by two orders of magnitude due to the extended interaction length~(Fig.~\ref{fig1}~b, middle panel). Additionally, the incorporation of strip lines into our photonic circuit offers a high degree of control over the terahertz electric field. The open-ended transmission line has a reflection coefficient exceeding  $90\%$, allowing us to realize a strip line terahertz cavity. Routing optical signals into the cavity provides means to generate terahertz modes at discrete frequencies that can be outcoupled into the farfield by an antenna~(Fig.~\ref{fig1}~b, right panel). Examples of practical applications that may adopt these concepts include enhancing the speed and resolution of millimeter-wave radar systems~\cite{zhu2023integrated}, implementation of Kerr combs for terahertz communications~\cite{shin_photonic_2023}, or terahertz imaging using photonics~\cite{li_plasmonic_2024}. This integration not only aligns with modern fabrication standards but also opens new avenues for chip-scale applications in telecommunications, spectroscopy, and computing.

\section*{Results}
\subsection*{Broadband photonics-integrated terahertz emitters}
\label{subsec_1}
Terahertz generation occurs when a short pulse of light undergoes optical rectification (OR) as it travels through a medium with a non-vanishing second-order nonlinearity $\mathrm{\chi^{(2)}}$. This nonlinear process leads to the generation of a second-order polarisation $\mathrm{P^{(2)}(t)}$ following the intensity envelope of the optical pulse $\mathrm{I(t)}$, which acts as a source of the emitted terahertz radiation. Generally, the bandwidth of the generated terahertz field is limited by the temporal duration of the optical pulses, while its amplitude scales linearly with the input pulse's intensity. In our case, the terahertz wave is generated inside the TFLN waveguide and it propagates along a transmission line (realized with gold strip lines) defined along the waveguide (see inset Fig.~\ref{fig1}~d). The exact geometry is detailed in the Supplementary Information Sec.~1~A. The generated terahertz pulse depends on three factors. First,  femtosecond short optical pulses are needed to maximize the intensity envelope, thereby enhancing the generation efficiency and the bandwidth. This is typically complicated by the presence of dispersive components in the pulse delivery setup to the terahertz emitter. Second, it is essential to fulfil phase matching between the terahertz and optical waves over the entire bandwidth offered by the femtosecond pulse. This would achieve maximal terahertz bandwidth. Finally, we need to outcouple the broadband terahertz wave into the farfield. 

\begin{figure}[h!]
    \centering
    \includegraphics[width=18cm]{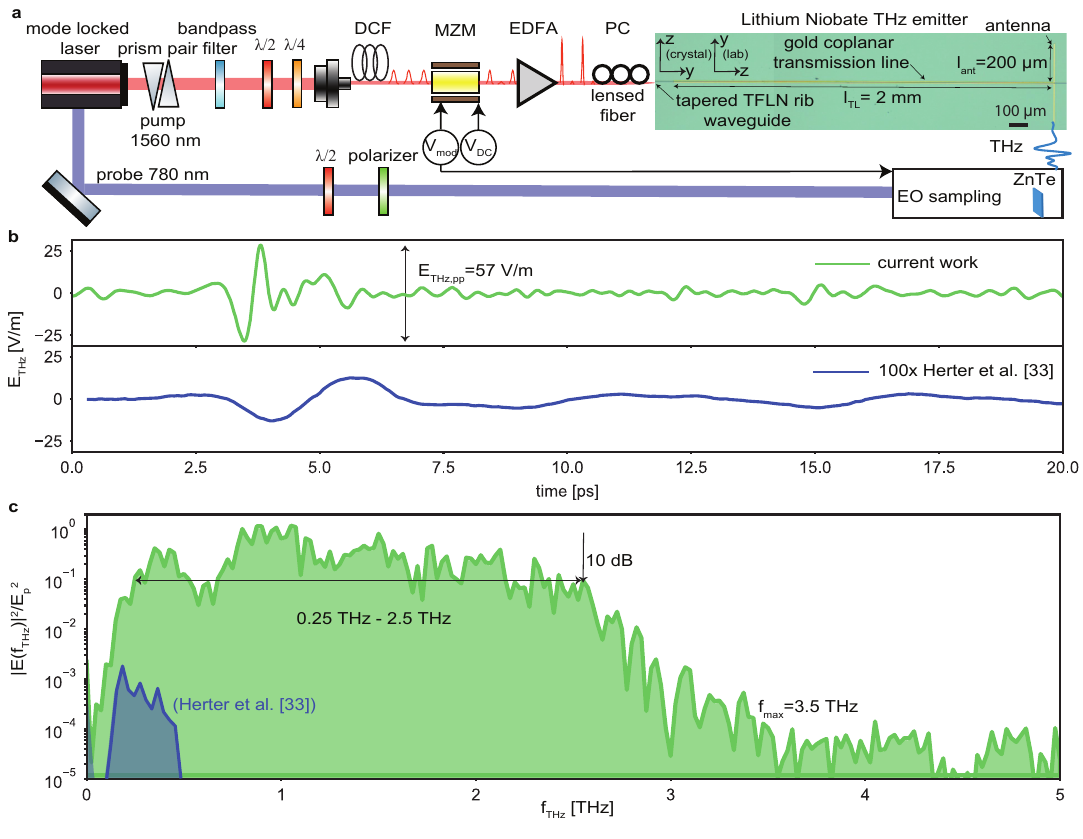}
    \renewcommand{\baselinestretch}{1} 
    \caption{\textbf{Photonics-integrated broadband terahertz emitter.} \textbf{a} Optical setup used for the characterization of the terahertz emitters. Modulated optical pump pulses are coupled via tapered fibers into the rib waveguide. Dispersion compensation in fiber ensures their compression at the input of the chip. A commercial electro-optic modulator controls the intensity of these pulses by means of applied bias voltages $\mathrm{V_{DC}}$ and $\mathrm{V_{mod}}$. Broadband terahertz radiation is generated inside the transmission line, and gets emitted into the farfield by terminating the structure with a broadband multi-wave antenna. This is detected using free-space electro-optic sampling inside a zinc telluride crystal. \textbf{b} Electric field of the TFLN emitter, featuring a transmission line of length $\mathrm{l_{TL} = 32~mm}$, terminated by a dipolar antenna of length $\mathrm{l_{ant}}$ on one end, as-measured inside a $\mathrm{1~mm}$ ZnTe crystal in comparison with current state of the art~($100\times$, lower panel). \textbf{c} The corresponding spectra is retrieved by taking the Fourier transform of the time traces, showing a dynamic range of 50~dB in intensity, and a maximum emission frequency of $\mathrm{f_{max} = 3.5~THz}$, clearly outperforming current our previous work~\cite{herter2023}. Both spectra are normalized to the pump pulse energy on-chip, which in our case is 100 pJ. EO = electro-optic, MZM = Mach-Zehnder modulator, EDFA = erbium doped fiber amplifier, PC = polarisation controller, TFLN = thin film lithium niobate, ZnTe = zinc telluride, DR = dynamic range, DCF = dispersion compensating fiber.}
    \renewcommand{\baselinestretch}{1.5} 
    \label{fig2}   
\end{figure}

To accomplish an efficient delivery of ultra-short pulses generated by a mode-locked laser, centered at wavelength of 1550 nm  (repetition rate of 100 MHz), we use a edge-coupling (butt-coupling) approach~(Fig.~\ref{fig2}~a). Furthermore, the dispersion is controlled using a prism pair and dispersion compensated fiber (DCF). To further mitigate the losses of the various components and achieve short pulses at the chip by self-phase modulation inside the fiber, an Erbium-doped fiber amplifier (EDFA) is employed before coupling to the chip. Our dispersion compensated setup generates pulses as short as 60~fs, corresponding to a 10 dB optical bandwidth of 75 nm (see Fig.~5~a and b in Supplementary Information). In contrast to previously used grating couplers~\cite{herter2023,Benea-Chelmus:20,Salamin2019}, edge couplers allow for coupling the entire bandwidth of the optical pump into the TFLN waveguides (see comparison of edge coupling and grating couplers in Fig.~5~c in Supplementary Information). To benefit from the largest component of the nonlinear tensor $d_{33}$, we align the polarization of the optical pulse to the z-direction of the TFLN.

To achieve phase matching between the terahertz and optical fields, we match the group velocity of the optical wave with the phase velocity of the terahertz fields by engineering the geometry of the transmission line, similar to the design concept used in broadband electro-optic modulators. The partial extension of the guided terahertz mode outside the LN material lowers its effective refractive index to $\mathrm{n_{TL} = 2.25-2.3}$ thereby matching it with the that of the pump (experimentally determined to be $\mathrm{n_g = 2.24}$, see Supplementary Information Sec.~1~B). Phase-matching leads to a linear increase of the terahertz field over the entire length of the transmission line $\mathrm{l_{TL}}$.  In the frequency domain, the terahertz electric field at the end of the transmission line is given by:
\begin{equation}
\label{OR}
    \mathrm{
     \tilde{E}_{THz}(l_{TL}, \omega_{THz}) = \frac{i \pi \chi^{(2)}E_0^2 \omega_{THz}^2 \tau^2 l_{TL} \Gamma_{overlap}}{4 c_0 n_{TL}(\omega_{THz}) sinh( \pi \omega_{THz} \tau / 2)} G_{TL}(\omega_{THz}) \cdot e^{-i \frac{\omega_{THz} l_{TL}}{c_0} n_{TL}(\omega_{THz})} 
     }
\end{equation}
where $\mathrm{E_0}$ is the pump electric field amplitude, $\tau$ is the FWHM of the pump pulse, $c_0$ is the speed of light in vacuum, $\mathrm{\chi^{(2)} = \chi^{(2)}_{333}=360~pm/V}$ is the second-order susceptibility along which all fields are oriented, and $\mathrm{n_g}$ is the pump group index at the central pump wavelength (Supplementary Information Sec.~2~A). The optical losses are negligible over the transmission line lengths we consider here. We introduce the phase-matching function:
\begin{equation}
\mathrm{G_{TL}(\omega_{THz}) = \frac{e^{-i \Delta k l_{TL}} - e^{-\alpha l_{TL} / 2 }}{i\Delta k l_{TL} - \alpha l_{TL} / 2}}
\end{equation}
 with $\mathrm{\Delta k = \frac{\omega_{THz}}{c_0}(n_g - n_{TL}(\omega_{THz}))}$ being the wave vectors mismatch and $\mathrm{\alpha}$ the loss coefficient of the guided terahertz field, defined as $\mathrm{E_{THz}(z) = E_{THz}(0) e^{-\alpha z/2 }}$. The coherence length of our transmission line is $\mathrm{5~mm}$ at $\mathrm{1~THz}$ and $\mathrm{0.3~mm}$ at $\mathrm{4~THz}$ (Fig~\ref{fig1}~d). 
The amount of generated terahertz electric field that effectively couples to the transmission line is also dependent on the overlap factor:
 \begin{equation}
  \mathrm{  {\Gamma_{overlap} = \frac{\iint_{(x,y)} g^2_{opt}(x,y) g^{*}_{THz}(x,y) dx dy}{\iint_{(x,y)} |g_{THz}(x,y)|^2 dx dy} }  }  
 \end{equation}
where $\mathrm{g_{THz}(x,y)}$ and $\mathrm{g_{opt}(x,y)}$ are the normalized spatial distributions of the THz and optical modes as visualized in the inset of \ref{fig1}~c, respectively. Using finite element method simulations (CST Microwave Studio), we find that the overlap $\mathrm{\Gamma_{overlap} = 0.1}$ depends weakly on frequency in the entire bandwidth spanning from 100~GHz to 4~THz (see Supplementary Information Fig.~7~c).

We note that similar to the RF loss in modulators, the terahertz loss of the transmission line and the terahertz-optical mode overlap also limit the bandwidth of the generated spectrum. In particular, both the radiative loss $\mathrm{\alpha_{rad}}$ (leaking of the terahertz wave through the substrate) and the absorption loss $\mathrm{\alpha_{abs}}$ (where terahertz field is absorbed by both gold and TFLN) impact the generation efficiency at higher frequencies more severely~(Supplementary Fig.~6). 
These loss values and the phase matching condition can all be optimized by tuning the dimensions of the transmission line (Supplementary Information Sec.~2~B). Following the full analysis, we find that $\mathrm{l_{TL} = 2~\mathrm{mm}}$ provides phase-matching for a broadband emission frequency up to $\mathrm{\omega_{THz} = 2\pi\times 3.5}$~THz. We terminate the transmission line with an antenna designed for a broadband and flat response.

We fabricate terahertz emitters with various lengths of the transmission line to investigate experimentally the effect of phase matching on the properties of the generated terahertz field. 
We use the electro-optic sampling technique to recover the temporal waveform as emitted from a dipolar antenna with $\mathrm{l_{ant}=200~\mu m}$ and width $\mathrm{w_{ant} = 5~\mu m}$ patterned at the end of the transmission line (a description of the entire experimental setup is given in Supplementary Information Sec.~1~C). We find that these antenna dimensions support emitting a broad terahertz spectrum by operating on the various higher-order modes (Supplementary Information Sec.~2~F). The measured terahertz signal for a transmission line of length $\mathrm{l_{TL} = 2~\mathrm{mm}}$ is shown in Fig.~\ref{fig2}~b, featuring an electric field amplitude of $\sim \mathrm{E_\text{THz,pp} = 57~V/m}$ (as measured) showing a $\sim$100 fold improvement in field amplitude over our previous work \cite{herter2023} and a significantly reduced duration of the terahertz pulse. 

The frequency spectrum is obtained by taking the Fourier transform of the temporal waveform revealing a broad and relatively flat-top emission, approaching a maximal frequency of 3.5 THz (10~dB bandwidth of approximately 2.5~THz). The maximal frequency of 3.5~THz is in good agreement with the calculated coherence length. Compared to our previous work \cite{herter2023}, we recorded a four-octave spanning spectrum with a dynamic range of approximately 50~dB in intensity, which makes our platform suitable for applications in spectroscopy requiring high signal-to-noise ratio. We experimentally compare various transmission line lengths of $\mathrm{l_{TL} = 0.12,~0.5~and~2~\mathrm{mm}}$ and find that a long transmission line benefits the generation efficiency of low terahertz frequencies compared to shorter transmission line (Supplementary Information Sec.~3~C).

We note that outside the transmission line, the strong terahertz-optical mismatch supports efficient generation only at a an angle of $\theta_c = 42^\circ$ with the yz-plane into the high-refractive index silicon substrate (see CST simulations Supplementary Information Sec.~2~E). By accounting for reflections (Fresnel and total internal reflection at angles beyond the critical angle) at the high-resistivity silicon substrate, at the detection crystal, limited numerical aperture of our detection system and asymmetric emission into the dielectric below and above the chip, we estimate the terahertz electric field at the end of the transmission line to be approximately a factor of $10^4-10^5$ higher than the measured one and amount to $\mathrm{E_\text{THz,TL} = 10^5-10^6~V/m}$ (see  Supplementary Information Sec.~2~F and 3~A). 
This agrees with our full-field model which accounts for all losses, phase matching, and spatial overlap of the mixing fields (see estimated fields in Supplementary Information Fig.~9).

\subsection*{Electro-optic control of the terahertz field amplitude}

Next generation of terahertz devices would require a precise control over the amplitude of the generated field, for example for amplitude-modulated terahertz communications. Equation~\ref{OR} reveals the numerous knobs that could be used, such as changing the nonlinear crystal thickness (interaction length), changing the input pulse duration, or the overlap between the terahertz and optical field. Among these possibilities,  controlling the amplitude of the input femtosecond pulse is the most widely used technique, typically implemented using a mechanical chopper. Chopping the input beam also aids the lock-in detection scheme towards recording the terahertz waveform. However, the slow speed of mechanical choppers ($\sim$kHz rates) compromises the signal to noise ratio (SNR) since flicker noise scales with $\propto 1/f$. More importantly, this technique provides only an on-off modulation without the ability to continuously control the amplitude of the input pulse. Even such an on-off modulation exhibits unfavorable performance due to slow rise and fall times, consequently leading to a worsened signal-to-noise ratio.
To achieve full control over the generated terahertz amplitude without these issues, we implement a fiber-pigtailed Mach-Zehnder electro-optic (EO) intensity modulator before our chip.

\begin{figure}[th]
    \centering
    \includegraphics[width=15cm]{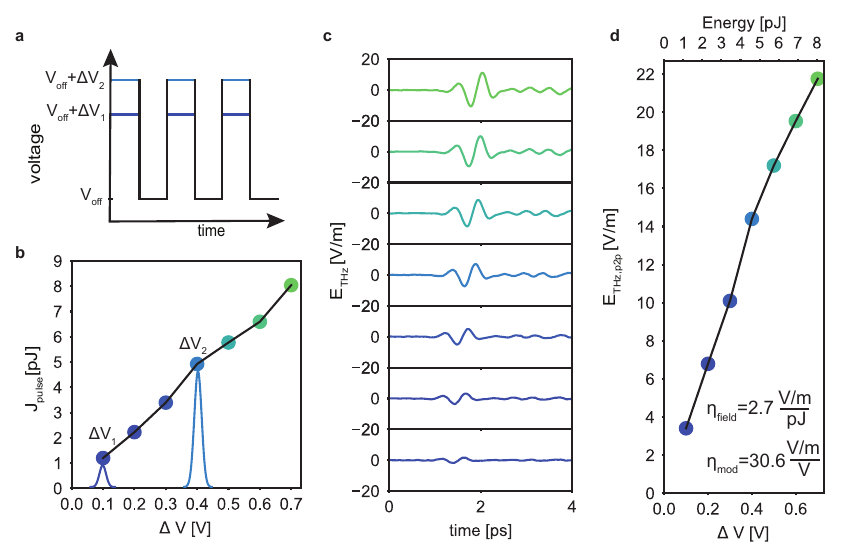} 
    \renewcommand{\baselinestretch}{1} 
    \caption{\textbf{Electro-optic control of the terahertz field amplitude.} \textbf{a}  The pump intensity modulation scheme used for THz amplitude control. The pump is switched between an off-state at $\mathrm{V_{off}}$, where no pump pulses reach the lithium niobate terahertz emitter, and an on-state $\mathrm{V_{off}+\Delta V}$ with controllable pump pulse energy.  In the case of our commercial Mach-Zehnder modulator the off-state voltage is $\mathrm{V_{off} = 6~V}$.  \textbf{b} Dependence of the pump pulse energies inside the TFLN chip on the applied modulation voltage $\mathrm{\Delta V}$ shows that energies up to $\mathrm{J_{pulse} = 8~pJ}$ depend linearly on applied voltages up to $\mathrm{\Delta V = 0.7~V}$. \textbf{c} measured terahertz electric fields for the various control voltages $\mathrm{\Delta V}$. \textbf{d} Peak-to-peak values of the terahertz electric field $\mathrm{E_{THz,pp}}$ exhibit linear dependence on $\mathrm{\Delta V}$. We measure a field generation efficiency of $\mathrm{\eta_{field} = \frac{E_{THZ,pp}}{J_{pulse}}=2.7\frac{V/m}{pJ}}$ and a field modulation efficiency of $\mathrm{\eta_{mod} =\frac{E_{THZ,pp}}{\Delta V}=30.6\frac{V/m}{V}}$.}
    \renewcommand{\baselinestretch}{1.5} 
    \label{fig3}  
\end{figure}
 
We adopt this EO control on a shorter device compared to the one in Fig.~\ref{fig2} ( $\mathrm{l_{TL} = 120~\mu m}$ vs $\mathrm{l_{TL} = 2~mm}$) which features a temporal profile closer to a single cycle. We operate the modulator by applying a square wave voltage at 1 MHz (below the repetition rate of the laser), toggling between an off-state and an on-state where a desired portion of the pump pulse energy is sent into the chip (Fig.~\ref{fig3}~a). By adjusting the amplitude of the ON state ($\mathrm{\Delta V}$), we send various input pulse energies $\mathrm{J_{pulse}}$ (Fig.~\ref{fig3}~b) resulting in a range of electric-field amplitudes (Fig.~\ref{fig3}~c). Consequently, we find that the corresponding peak-peak amplitude $\mathrm{E_{THz, pp}}$ of the emitted terahertz electric field exhibits a linear dependence on the modulated pump pulse energy and, consequently, on the applied voltage $\mathrm{\Delta V}$ (Fig. \ref{fig3}~d). Analysing the slope, we find a generation efficiency of $\mathrm{\eta_{field} = \frac{E_{THZ,pp}}{J_{pulse}}=2.7\frac{V/m}{pJ}}$ and a field modulation efficiency of $\mathrm{\eta_{mod} =\frac{E_{THZ,pp}}{\Delta V}=30.6\frac{V/m}{V}}$. 
We further analyze the noise properties of our all-optical terahertz emission and detection setup. We find that for a modulation frequency of 1~MHz, the noise of our lock-in detection is only slightly above the shot-noise limit (the measured voltage fluctuation is $\mathrm{\delta V_{meas} = 160~nV}$ while the expected shot-noise limited voltage fluctuation is $\mathrm{\delta V_{shot~noise} = 148~nV}$, see Supplementary Information Sec.~3~B.  This experimentally confirms the advantages of MHz-speed electro-optic modulation of the input pulse for reaching low noise at the fundamental shot noise limit.

\subsection*{Terahertz cavities from coplanar transmission lines}

\begin{figure}[h!]
    \centering
    \includegraphics[width=16cm]{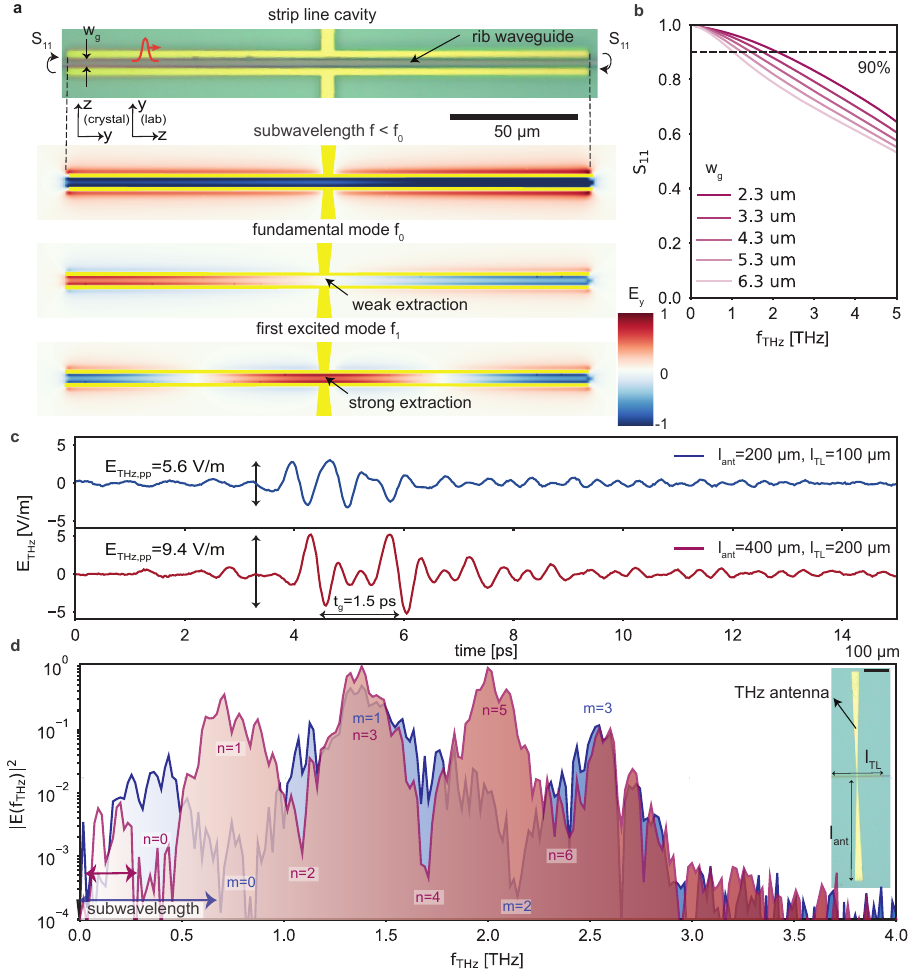}
    \renewcommand{\baselinestretch}{1} 
     \caption{\textbf{Strip line terahertz cavities from transmission lines.} \textbf{a} A terahertz cavity is formed by terminating a transmission line with an open end on both sides. We excite the cavity modes by optical rectification of pump pulses that travel along the rib waveguide. To access these cavity modes, we pattern a broadband antenna centrally around the transmission line. \textbf{b} The reflection coefficients of the open termination reaching above $\mathrm{S_{11}> 90~\%}$ at frequencies below 1.7~THz and $\mathrm{S_{11}= 78~\%}$ at 3~THz, from CST simulation. These values depend on the sub-wavelength dimension of the cross-section of the transmission line. \textbf{c} Experimental results of the terahertz field emitted from different strip line cavities with length of $\mathrm{l_{TL} = 100~\mu m}$ and  $\mathrm{l_{TL} = 200~\mu m}$. Clearly visible, especially in the case of the longer cavity, is a train of pulses which in that case exhibits a group delay of $\mathrm{t_g = 1.5~ps}$. The THz field amplitude of the initial terahertz pulse is approximately twice larger for the case of the twice longer strip line cavity ($\mathrm{E_{THz,pp,1} = 9.4~V/m}$ vs $\mathrm{E_{THz,pp,1} = 5.6~V/m }$), in line with expectation due to the ratio of the interaction lengths.  \textbf{d} The Fourier transforms of the terahertz electric field reveal broadband spectra that feature discrete modes, experimentally confirming the build-up of cavity modes. Indeed, the twice longer strip line cavity features twice as many modes compared to the shorter one with resonances that align well in frequency.}
    \renewcommand{\baselinestretch}{1.5} 
    \label{fig4}   
\end{figure}

Apart from the transmission lines, integrated cavities are crucial to expanding the capabilities of terahertz integrated photonic devices. They serve important functionalities such as temporarily storing light, spectral filtering, and field enhancement dictated by the their facet reflectivity~\cite{doi:10.1142/5485}. One way to realize a strip-line cavity would be to leave both sides of a transmission line open-ended (unterminated), which would constitute a strip line cavity~\cite{PozarBook}. A given cavity mode would either have a node (for odd modes) or an anti-node (for even modes) at the center of the cavity strip line. 
Therefore, an antenna placed at the center can extract the odd modes into the far field (strong extraction), while not efficiently out-coupling odd modes into the far-field (weak extraction) (Fig.~\ref{fig4}~a).
Since the transverse dimensions of our transmission lines are highly sub-wavelength compared the generated terahertz wavelength, the reflection coefficient exceeds $\mathrm{S_{11} = 79\%}$ in the entire band up to $\mathrm{f_{THz}} =  3~\mathrm{THz}$ (see the $\mathrm{S_{11}}$ in Fig.~\ref{fig4}~b). 
We fabricate two strip-line cavities with lengths of $\mathrm{l_{TL}=100~\mu m}$ and $\mathrm{l_{TL}=200~\mu m}$ to have different mode profile at the antenna. We selected these two candidate geometries to evaluate the temporal and spectral behavior of the generated field (circulating inside the cavity) with respect to strip-line length and antenna dimensions (Fig.~\ref{fig4}~c).

As the optical pump propagates inside the transmission line and the terahertz radiation accumulates, a portion of the terahertz field is coupled to the antenna, leading to the initial terahertz pulse formation. The remaining terahertz field continues to build up until it reaches the end of the transmission line and reflects back. Once it reaches the antenna again, it is coupled out with a group delay of $\mathrm{t_g = \frac{n_g l_{TL}}{c}}$ with respect to the initial terahertz pulse. 
This cycle repeats until the terahertz field has been entirely coupled out into the far field, absorbed in the lossy materials, or leaked out through the cavity mirrors.
The build-up in the shorter transmission line is slightly weaker in terahertz field ($\mathrm{E_{THz,pp,1} = 5.6~V/m}$) compared to the longer one ($\mathrm{E_{THz,pp,2} = 9.8~V/m}$) due to shorter interaction length. However, its group delay $\mathrm{t_g \sim 0.7~ps}$ is comparable to the duration of the terahertz pulse, in contrast to the case of the longer cavity. Consequently, the terahertz field is stored inside the cavity for shorter times (approximately 4~ps) as opposed to the longer cavity, where it exceeds 6~ps, showcasing how cavities can extend the light lifetime inside a cavity. 

The Fourier Transforms of these time-domain traces reveals formation of discrete modes with their exact frequency depending on the length of the transmission line (Fig.~\ref{fig4}~d), unlike the broadband spectrum of the travelling wave emitter in Fig.~\ref{fig2}~c. The spectral analysis of the terahertz emission reveals two key findings. Firstly, the fundamental cavity mode (n = 0) is not extracted by the antenna due to its placement at the minima of the cavity modes. Secondly, at frequencies below the fundamental mode, the antenna operates in the Hertzian dipole regime with sub-wavelength interaction length (which was explored in Ref.~\cite{herter2023}). This regime is characterized by a uniform distribution of the terahertz field showing no visible z-dependency (Fig.~\ref{fig4}~a, sub-wavelength $\mathrm{l_{int}<\frac{\lambda_{THz}}{2}}$). Our full cavity model captures faithfully the experimental details such as field amplitudes difference in the time domain, and spacing of the measured resonances in the frequency domain (full details in Supplementary Information Sec.~2~C).

\section*{Discussion}\label{Conclusion}
In summary, we demonstrate the hybrid integration of terahertz transmission lines with photonic integrated circuits enabling broadband terahertz generation up to 3.5~THz, spanning four octaves. We achieve this by overcoming the significant limitation of phase mismatch between the optical and terahertz signals, which previously caused their early walk-off in bulk. Similar to RF equivalents, our phase-matched terahertz transmission lines confine the terahertz electric field as it builds up and guide it along with the optical pump, enabling longer interaction lengths beyond a few millimeters and supporting close to single-cycle terahertz pulses with amplitudes two orders of magnitude larger than state-of-the-art. We read out these terahertz electric fields all-optically, and demonstrate a detection close to shot-noise limit with a dynamic range exceeding 50~dB in intensity. Our phase-matching at telecommunication wavelengths is particularly significant, as there are no alternative bulk generation crystals that are phase matched at 1550 nm other than organic crystals~\cite{Vicario:15}. Most generation crystals only work at other wavelengths, such as 780~nm\cite{vidal_impact_2011} or 1050~nm~\cite{aoki_broadband_2017}. This advantage opens doors for a wide variety of optical communication infrastructure to be implemented in our system. As an example, we use a fiber-based EO-amplitude modulator to control the amplitude of the terahertz emission. 
This not only provides future opportunities to integrate more functionalities on TFLN chips, but also showcases a path towards all-optical terahertz signal processing with gigahertz analog bandwidth. %facilitating 6G systems~\cite{akyildiz20206g}. 
Combining the terahertz transmission lines with impedance-matching antennas for off-chip emission, additionally addresses free-space terahertz applications. Furthermore, we demonstrate the first proof-of-principle photonics-integrated terahertz strip-line cavities. We experimentally verify that efficient generation of specific modes is feasible by choosing the strip line length and placing a terahertz antenna at specific locations. Such a toolkit could provide a versatile way for tailoring terahertz emission for a specific selection of frequencies \cite{shams2022thin}. Finally, our system offers unique control over the terahertz radiative loss in the transmission lines through tweaking its dimensions and lower terahertz material loss because of thin layer of LN, providing access to a low-loss regime required for building more complex terahertz circuits. 

Our concept can be applied to continuous-wave applications requiring narrow linewidth terahertz signals by replacing the input pulse with two continuous-wave tones separated by a desired terahertz frequency~\cite{Wang:08}. Such all-optical generation of narrow-band terahertz tones can complement current all-electronic methods such as multiplier chains or quantum cascade lasers (QCL)~\cite{bosco_patch-array_2016}. This is beneficial in the context of terahertz-speed modulators for optical fiber communications by reducing the reliance on external instrumentation~\cite{Zhang:22,alloatti2014100} with significant ohmic loss. Additionally, local oscillators in heterodyne terahertz spectroscopy would benefit from all-optical narrow-band terahertz tones without requiring cryogenic cooling, as seen with QCLs. Their capability to generate terahertz radiation above 1~THz could allow measuring hyperfine transitions~\cite{DROUIN20112} e.g. of  [CII] at 158~$\mathrm{\mu m}$~\cite{refId0}, [NII] at 122~$\mathrm{\mu m}$ and 205~$\mathrm{\mu m}$, Al above 2 THz. More importantly, their compatibility with in-situ tuning of the optical signals could provide the fine- and broad tuning needed for high-resolution planetary heterodyne spectroscopy.  

The concept of phase matching using transmission lines can be readily applied to optimize the efficiency of other nonlinear processes enabling for example terahertz detection in the same platform. Providing sub-wavelength confinement of terahertz waves, our phase-matched strip line cavities also facilitate quantum electrodynamics experiments. They provide vacuum fields that are three orders of magnitude stronger than those in free space experiments~\cite{benea-chelmusElectricFieldCorrelation2019}, simplifying their direct detection by routing optical waveguides into these cavities. Finally, the combination of strongly enhanced local terahertz field inside the transmission lines, terahertz dispersion engineering (by means of designing the transmission line geometry) and extremely high nonlinear coefficient of TFLN ($\mathrm{d_{33} = 5870~pm/V}$~\cite{PhysRevB.7.5345} at terahertz wavelengths) can enable all-terahertz mixing (similar to all-optical mixing). These, combined with the mature wafer-scale fabrication of TFLN~\cite{Luke:20} allows for integration of numerous photonic components such as femtosecond pulse sources \cite{yu2022integrated, wang_integrated_2024}, electro-optic frequency combs~\cite{hu_high-efficiency_2022}, or self injection-locked laser sources with our terahertz circuits~\cite{ulanov2024synthetic}. We anticipate that the design guidelines we propose will become crucial in future terahertz applications both on- and off-chip, such as free-space communications, terahertz-speed computing, data interconnects, ranging and metrology. 

\section*{Methods}
\subsection*{Fabrication}\label{Meth:fab} 
The chips are fabricated on 600 nm of X-cut lithium niobate, bonded to 4700 nm of thermally grown oxide on an approximately 500 $\mathrm{\mu m}$-thick double-side polished high-resistivity silicon substrate. The waveguides are patterned using electron-beam lithography (Eliionix ELS-HS50) and Ma-N resist. These waveguides are then etched into the LN layer using  $\mathrm{Ar^+}$ ions, followed by annealing in an $\mathrm{O_2}$ environment to recover implantation damages and improve the absorption loss of the platform \cite{shams2022reduced}. Subsequently, the devices are clad with 800 nm of Inductively Coupled Plasma - Chemical Vapor Deposition (Oxford Cobra), followed by another annealing. The electrodes are defined using a self-aligned process, including patterning with PMMA resist, dry etching, and lift-off using the same resist. The electrodes are deposited using electron-beam evaporation (Denton) with 15 nm of Ti and 300 nm of Au.

\subsection*{Optical setup}\label{Meth:setup}
The mode-locked laser used for THz generation and detection is C-Fiber 780 from Menlo Systems. it provides 100 MHz repition rate with average power values up to 500 mW and pulse duration in the range of 60 fs. The 1560 nm output is used as the pump for THz generation and the 780 nm output is used as the probe for electro-optic sampling. Before coupling the pump laser into fiber, a Silicon prism pair (SIFZPRISM25.4-BREW from Crystran) is used in a double pass configuration to provide tunability of group velocity dispersion and the spectrum is filtered using a bandpass filter with FWHM of 12 nm at 1530 nm (FBH1530-12 from Thorlabs). To precompensate the total dispersion from the 11 m of single mode fiber inside the setup, we use dispersion compensating fiber (2x PMDCFA5 from Thorlabs). Modulation of the pump is done using the component MXAN-LN-10-PD-P-P-FA-FA from IX-Blue, which can support up to 10 GHz. Losses from coupling to fiber and form the intensity modulator are compensated using the optical amplifier EDFA100S from Thorlabs which operates at maximum pump current. After amplification, the high peak power in fiber triggers spectral broadening (from 12 nm to 50 nm) and Raman scattering leads to shift in spectrum toward longer wavelength (supplementary material). The broad spectrum leads to efficient optical rectification inside $\text{LiNbO}_3$. The polarization is adjusted before coupling to the chip using a polarization controller to couple into the TE mode of the x-cut $\text{LiNbO}_3$ waveguide and exploit $d_{33}$ coefficient. Finally, a 1 m bare lensed fiber with 5 $\mu m$ minimum spotsize is used for edge coupling and the coupling efficiency from fiber to the chip reaches 13\%. For measuring the signal using the convential electro-optic sampling technique, a ZnTe and GaP crystals are used. The pulse length of the 780 nm probe 200 fs. Parabolic mirrors with 50 mm reflective focal length are used to collect and focus the THz radiation. The probe signal is measured using a balanced photodetector (PDB465A from Thorlabs) UHFLI 600 MHz Lock-in Amplifier with time constants between 300 ms and 1 s. The delay stage scans the THz signal by moving the probe pulse in time with 25 fs steps.

\vspace{1cm}

\textbf{Data Availability}
The data generated in this study will be made publicly available in the Zenodo database prior to publication.

\medskip
\textbf{Code Availability}
The code used to plot the data within this paper is available in the Zenodo database prior to publication.

\section*{References}
\bibliographystyle{paper}
\bibliography{bibliography-main}

\medskip
\textbf{Acknowledgements}
Y.L., A.G. and I.C.B.C. acknowledge funding from the European Union’s Horizon Europe research and innovation programme under project MIRAQLS with grant agreement No 101070700. S.R., A.T. and I.C.B.C acknowledge funding from the Swiss National Science Foundation under PRIMA Grant No. 201547. S.R. acknowledges financial support from the Hans Eggenberger foundation (independent research grant 2022) and from the Swiss National Science Foundation (Postdoc.Mobility, grant number 214483). A.S.-A. and M.L. acknowledge funding from Defense Advanced Research Projects Agency (HR0011-20-C-0137). L. M. acknowledges Capes-Fulbright and Behring foundation fellowships. The fabrication of these chips was performed in part at the Center for Nanoscale Systems (CNS), a member of the National Nanotechnology Coordinated Infrastructure Network (NNCI), which is supported by the National Science Foundation under NSF Award no. 1541959.

\medskip
\textbf{Author contributions} I.C.B.C., Y.L and A.S.-A. conceptualised the project. Y.L. built the terahertz-optical setup and carried out the measurements. A.T. and S.R. assisted with building the optical setup. Y.L. and A.G. performed the CST simulations. S.R. assisted with the CST simulations of the antenna design. A.S.-A., S.R., and L.M. fabricated the devices. A.G. derived the theoretical description of the terahertz emission. Y.L., A.S.-A., A.G., and I.C.B.C wrote the manuscript with feedback from all authors. I.C.B.C.. and M.L. supervised this work.

\medskip
\textbf{Competing interests} A U.S. patent application US18/213,691 has been filed on the subject of this work by the Technology Transfer Office at EPFL. 

\medskip
\textbf{Disclaimer}
The views, opinions and/or findings
expressed are those of the author and should not be interpreted as representing the official views or policies of the Department of Defense or the U.S. Government.

\textbf{Corresponding authors} Correspondence to Yazan Lampert (yazan.lampert@epfl.ch), Amirhassan Shams-Ansari (ashamsansari@seas.harvard.edu) or Cristina Benea-Chelmus (cristina.benea@epfl.ch). 

%\listoffigures
\newpage
%\nolinenumbers
\pagenumbering{gobble}
\centering{\textbf{\Large Supplementary information for} \\  Photonics-integrated terahertz transmission lines}\\
\date{}
{\small Y. Lampert, A Shams-Ansari, A. Gaier, A. Tomasino, S. Rajabali, L. Magalhaes, M. Lončar, I.-C. Benea-Chelmus}

\newpage



\section*{Supplementary material}

# Contents

# 1 Details on TFLN samples and experimental setup

## A  Geometry and properties of the Lithium Niobate devices

Details on the geometry of the measured samples are given in Supplementary Fig. 1 and all dimensions are provided in Table 1. These dimensions are used for our CST simulation. The extraordinary refractive index of Lithium Niobate in the telecom range is calculated from [1] and imported to the material properties to determine the effective group refractive index for TE mode inside the X-cut Lithium Niobate rib waveguide. Similarly, the refractive index at THz frequencies is calculated using the double Lorentz model and its coefficients from [2]. Since Lithium Niobate is transparent in the telecom range, its absorption is neglected. In contrast, it is included for our THz simulations.

| | | |
|---|---|---|
| thickness of the Si substrate | $h_{Si}$ | 500 $\mu$m |
| thickness of the SiO$_2$ layer | $h_{SiO_2}$ | 4.7 $\mu$m |
| thickness of Thin-Film LiNbO$_3$ layer | $h_{TF}$ | 300 nm |
| thickness of the transmission line layer | $h_{TL}$ | 300 nm |
| thickness of the antenna layer | $h_{ant}$ | 300 nm |
| height of the LiNbO$_3$ waveguide | $h_{wg}$ | 600 nm |
| thickness of the SiO$_2$ cladding | $h_{cladding}$ | 1 $\mu$m |
| narrow width of the LiNbO$_3$ waveguide | $w_{wg,narrow}$ | 1.5 $\mu$m |
| width of the LiNbO$_3$ waveguide | $w_{wg,wide}$ | 1.85 $\mu$m |
| width of the SiO$_2$ cladding | $w_{cladding}$ | 3.3 $\mu$m |
| width of the antenna | $w_{ant}$ | 5 $\mu m$ |
| width of the gold layer inside the transmission line | $w_{TL}$ | 3.5 $\mu m$ |
| width of the gap between the two leads of the transmission line | $w_g$ | 3.3 $\mu m$ |
| length of the transmission line | $l_{TL}$ | 120 $\mu m$ |
| length of the antenna | $l_{ant}$ | 200 $\mu m$ |

**Supplementary Table 1:** Dimensions of the simulated and fabricated devices

## B  Measurement of the pump group index

We measured the transmission spectrum of the LiNbO$_3$ waveguide using the tunable continuous-wave laser (Keysight, 8164A) with a resolution of 0.1 pm. Due to the reflection at the facets of the chip, the modulation of transmission spectrum is observed (Fig. 2), which we attribute to the formation of the Fabry-Perot modes in the waveguide. Taking the difference between local maxima of the transmission, we retrieve the group index using the expression:

$$n_g = \frac{\lambda^2}{2 \cdot \Delta\lambda \cdot L} \qquad (1)$$

$\lambda$ is the wavelength, $\Delta\lambda$ is the difference between two maxima, and L is the length of the waveguide.

Averaging over the range 1495-1505 nm, we retrieved group index at optical frequency to be $n_g = 2.24 \pm 0.05$, in line with simulations shown in Fig. 1 of the main text.

## C  Experimental setup

The experimental setup is depicted in Supplementary Fig. 3. Detection is done with conventional electro-optic sampling [3]. A probe beam at $\lambda_{pr} = 780$ nm with pulse length of 200 fs is focused at the detection crystal and its

3

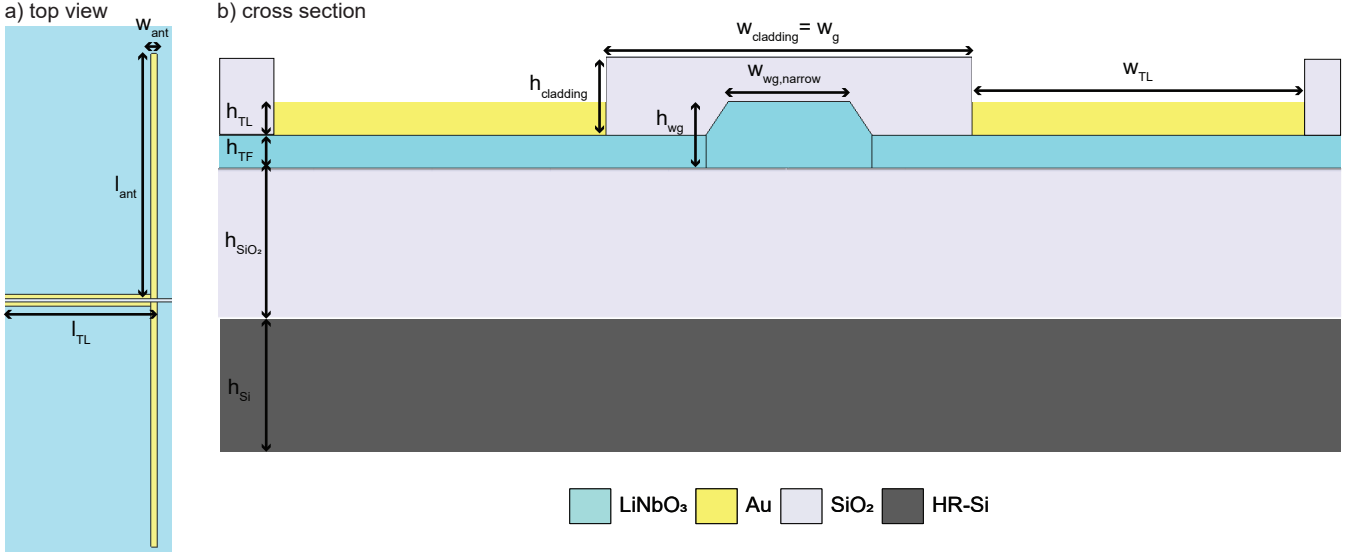

**Supplementary Fig. 1: Top view and cross section of the fabricated chip. a** top view of the device showing the transmission line with length $l_{int}$ terminated by a dipole antenna with length $l_{ant}$ and width $w_{ant}$. **b** cross section of the transmission line.

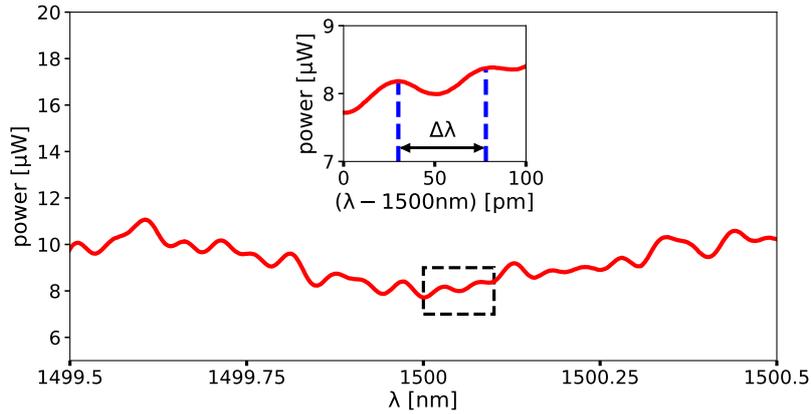

**Supplementary Fig. 2:** Transmission spectrum of the LiNbO$_3$ waveguide with length L = 10mm.

polarization is converted into circular at a subsequent quarter wave plate. The two orthogonal polarization components are split using a polarization beam splitter (Wollaston prism) and the difference in optical power is measured using balanced detection, thereby providing common-mode noise rejection. The terahertz beam induces birefrengence at the detection crystal which introduces a slight ellipticity in the otherwise circularly polarised probe beam that eventually leads to intensity imbalance at the balanced detector which is proportional to the incident terahertz electric field. The balanced detector has a bandwidth of 200 MHz which is larger than the 100 MHz repetition rate of the mode locked laser. Therefore, a low-pass with 80 MHz cut-off is used before feeding the voltage signal to the lock-in detector. ZnTe is used for electro-optic sampling of the emitted THz fields and its specifications are provided in table 2[4] where $\omega_{\text{pr}} = 2\pi \times \frac{c_0}{\lambda_{\text{pr}}}$ is the angular frequency of the probe beam. Both THz and probe are aligned along the y-axis of the lab which is parallel to the $[1\bar{1}0]$ axis of the detection crystals, enabling use of the nonlinear coefficient $r_{41}$ to detect the terahertz electric field.

We apply the following formula to retrieve the THz electric field in absolute values for the Fig. 2, 3 and 4 in the

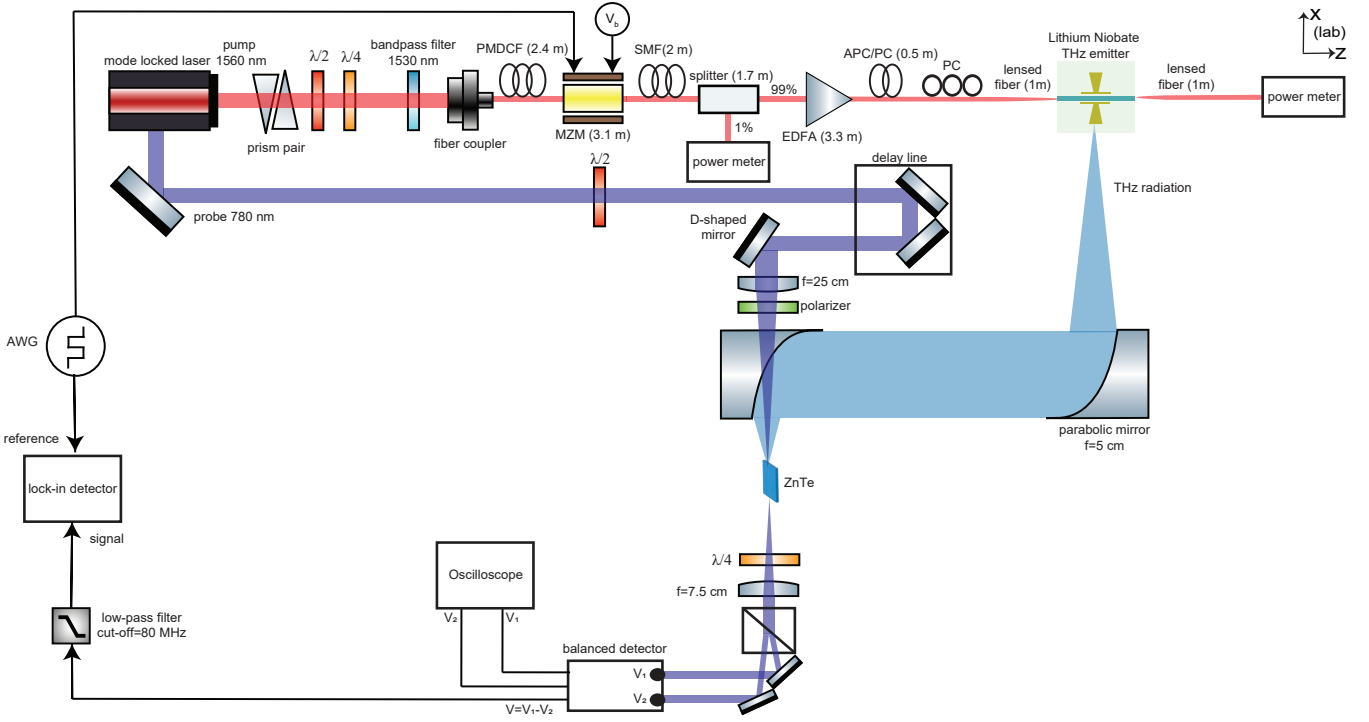

**Supplementary Fig. 3: Experimental setup for THz generation and detection.** PMDCF: polarization maintaining dispersion compensating fiber, SMF: single mode fiber, MZM: Mach-Zehnder-modulator, EDFA: Erbium-Doped Fiber Amplifier, APC: angle physical contact, PC: physical contact, AWG: arbitrary wave generator.

main text [5]

$$E_{det}(t) = \frac{c_0 \cdot \Delta I}{I_{pr} \cdot \omega_{pr} \cdot r_{41} \cdot L \cdot n(\omega_{pr})^3} \quad (2)$$

| crystal | thickness L | nonlinear coefficient | crystal orientation | $n(\omega_{THz})$ | $n(\omega_{pr})$ |
|---------|-------------|----------------------|---------------------|-------------------|------------------|
| ZnTe | 1 mm | $r_{41}$=3.9 pm/V | $[1\bar{1}0]$ | 3.4 | 2.87 |

**Supplementary Table 2:** Used crystal for electro-optic sampling and its specifications [4]

$\Delta I$ is the difference in detected intensity which is induced by the THz field. Experimentally, it is calculated using $\Delta I = \Delta V/G_{RF}$ where $\Delta V$ is the voltage difference between the two photodiodes at the RF output of the balanced detector and $G_{RF}$ is the conversion gain [V/W] of the photodetector. $P_{pr}$ is the total probe intensity. It is determined by reading out the voltage from the DC monitor of each photodiode and accounting for the DC signal gain G. This leads to the formula $I_{pr} = V_1/G + V_2/G$. To achieve balanced detection, the polarization of the probe is adjusted such that $V_1 = V_2$. Both detection crystals cover a broad range of THz radiation while using the same probe wavelength 780 nm. Their response can be summarized into a function $F(\omega_{pr}, \omega_{THz})$ which contains absorption, phase matching at a fixed crystal length $L$, strength of the nonlinearity and the Nyquist limit determined by the pulse length of the probe, providing a mean to compare the two crystals [3][6]:

$$F(\omega_{pr}, \omega_{THz}) = A_{opt}(\omega_{THz}) \cdot r_{41} \cdot \Delta\phi(\omega_{pr}, \omega_{THz}) \quad (3)$$

$A_{opt}(\omega_{THz})$ is the autocorrelation of the probe pulse which is estimated to have FWHM pulse length of 200 fs.

$$A_{opt}(\omega_{THz}) = \int_{-\infty}^{\infty} E_{pr}(\omega'_{pr} - \omega_{pr})^* E_{pr}(\omega'_{pr} - \omega_{pr} - \omega_{THz}) d\omega'_{pr} \quad (4)$$

$\Delta\phi(\omega_{\text{pr}}, \omega_{\text{THz}})$ is the phase matching function of the used crystal

$$\Delta\phi(\omega_{\text{Pr}}, \omega_{\text{THz}}) = \frac{e^{i\Delta kL} - 1}{i\Delta k} \quad (5)$$

$$\Delta k = \frac{\left(\omega_{\text{pr}} \cdot n(\omega_{\text{pr}}) + \omega_{\text{THz}} \cdot (n(\omega_{\text{THz}}) + i\kappa(\omega_{\text{THz}})) - (\omega_{\text{pr}} + \omega_{\text{THz}}) \cdot n(\omega_{\text{pr}} + \omega_{\text{THz}})\right)}{c_0} \quad (6)$$

Both absorption and dispersion have been included. The results of our calculations are shown for a ZnTe crystal of length L = 1 mm in Supplementary Fig. 4.

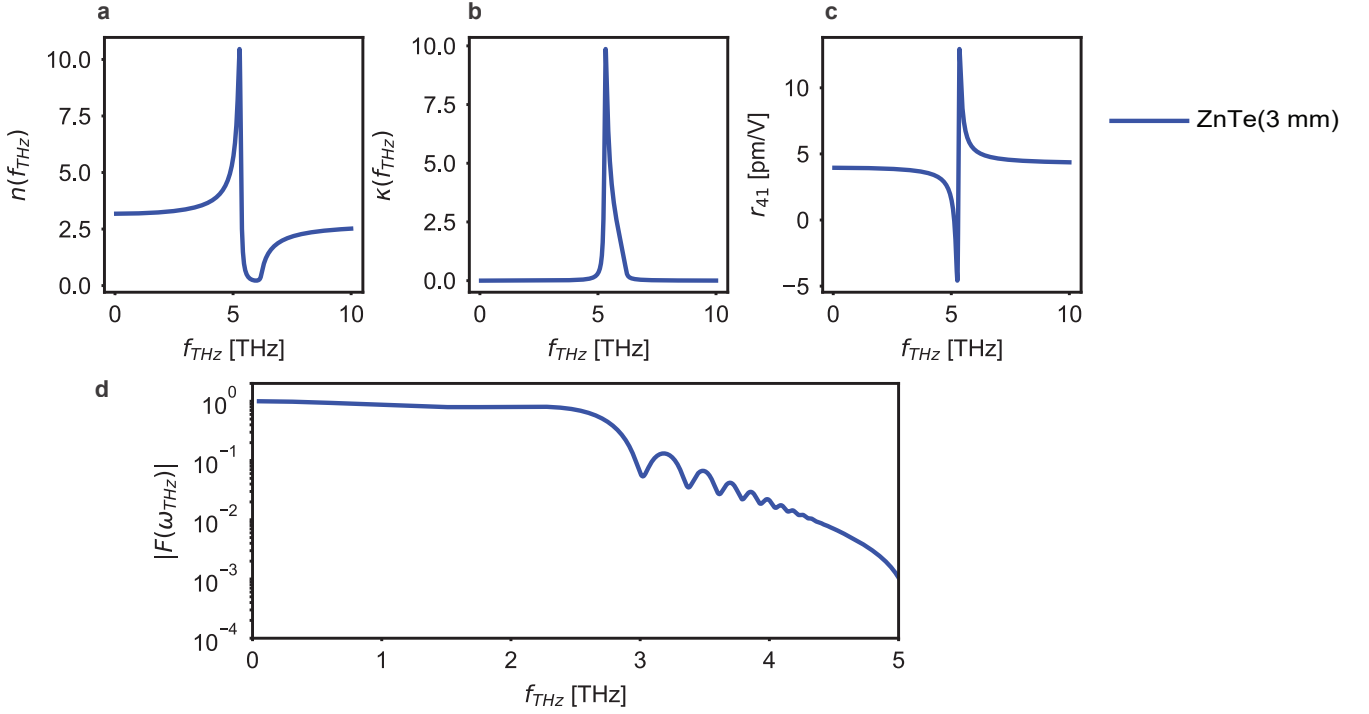

**Supplementary Fig. 4: Properties of ZnTe as a detection crystal**. **a** Real part of the refractive index. **b** Imaginary part of the refractive index. **c** Nonlinear electro-optic coefficient $r_{41}$. **d** Calculated response using equation 4 with probe wavelength of 780 nm.

To generate terahertz pulses, we use a telecom pump at 1550 nm which we characterize at the input of the chip using an autocorrelator. A hyperbolic sech fit provides a pulse length of 60 fs, illustrated in Supplementary Fig. 5 a. The optical spectrum at input of the chip is measured using an optical spectrum analyzer as shown in Supplementary Fig. 5 b. Spectral broadening is clearly observed after propagation in fiber and the spectrum at the chip's input spans over 100 nm. To confirm that broad spectra can be coupled into the Lithium Niobate devices, the transmission is characterized and the performance is compared to grating couplers. Supplementary Fig. 5 c shows the total transmission for the two facets as function of wavelength using grating couplers and edge coupling. In the edge coupling we implement, mode-matching between the waveguide and the lensed optical fiber is enhanced by adiabatically tapering the rib LiNbO$_3$ waveguide from a width of 1500 nm to 800 nm at the edge of the chip over a length of 3 mm. Edge-coupling allows for a broadband operation of the device accommodating for the entire bandwidth of femtosecond pulses.

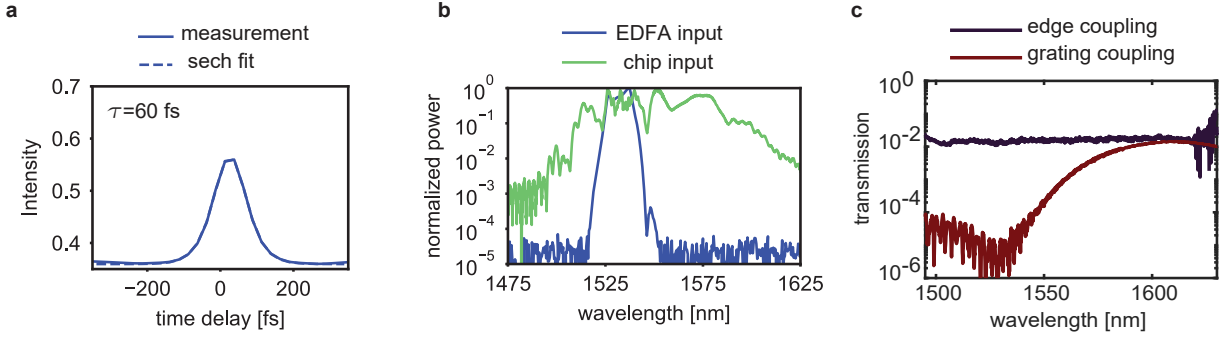

**Supplementary Fig. 5: Characterization of the optical pump**. **a** pulse intensity measurement using autocorrelation. **b** optical spectrum at input of the EDFA and input of the THz emitter. **c** Total transmission of the Lithium Niobate waveguide (for the two facets) using edge coupling compared to grating couplers.

## 2  Theory and modelling of chip-scale emitters

### A  Theory of terahertz generation inside photonics-integrated transmission lines

In this section, we derive the expressions for the THz electric field generated during the propagation of the femtosecond pulse in optical waveguide. Here, we follow the same procedure, which was developed in [7]. We consider the femtosecond pulses to propagate in the optical waveguide along z lab axis. Their electric field is linearly polarised (x component) and given by:

$$E_p(x,y,z,t) = \frac{1}{2} E_0 \varepsilon(z,t) g_{opt}(x,y) e^{i(\omega_p t - k_p z)} + \text{c.c.} \tag{7}$$

where $\omega_p$ is center carrying frequency, $k_p = 2\pi n_p/\lambda_p$ is the wave vector, $g_{opt}(x,y)$ is the optical mode profile, normalized such that $\iint_{(x,y)} |g_{opt}(x,y)|^2 dxdy = S_{eff}$, where $S_{eff}$ is the effective mode area, $\varepsilon(z,t)$ is an envelope slowly varying during the propagation, and $E_0$ is the pump pulse amplitude. The normalization is such, that the $\int_{-\infty}^{\infty} \iint_{(x,y)} 1/2 \varepsilon_0 c_0 \varepsilon_{eff} |E_p(x,y,z=0,t)|^2 dxdydt = J_{opt}$, where $J_{opt}$ is the pulse energy, and $c_0$ is the speed of light in vacuum. The terahertz wave is generated via difference frequency generation between frequency components belonging to the pulse. The evolution of the THz electric field is described by the non-linear wave equation:

$$\left[ \nabla^2 - \frac{1}{c_0^2} \frac{\partial^2}{\partial t^2} \right] E_{THz}(x,y,z,t) = \frac{1}{\varepsilon_0 c_0^2} \frac{\partial^2}{\partial t^2} \left( P_{THz}^L(x,y,z,t) + P_{THz}^{NL}(x,y,z,t) \right) \tag{8}$$

where $\varepsilon_0$ is the vacuum permittivity. The linear and non-linear polarizations, $P_{THz}^L(x,y,z,t)$ and $P_{THz}^{NL}(x,y,z,t)$ take the form of:

$$P_{THz}^L(x,y,z,t) = \varepsilon_0 \int_{-\infty}^{t} \chi^{(1)}(t-t') E_{THz}(x,y,z,t') dt' \tag{9}$$

$$P_{THz}^{NL}(x,y,z,t) = \varepsilon_0 \chi^{(2)} |E_p(x,y,z,t)|^2 \tag{10}$$

where $\chi^{(1)}$ is the complex linear susceptibility, and $\chi^{(2)} \approx 360 \text{pm/V}$ is the second-order susceptibility of LiNbO$_3$ at microwave and terahertz frequencies (below the phonon lines) [8, 9, 10]. Taking the Fourier transformation of 8, taking into account the relation between linear polarization and electric field $\tilde{P}_{THz}^L(x,y,z,t) = \varepsilon_0 \chi^{(1)}(\omega_{THz}) \tilde{E}_{THz}(x,y,z,\omega_{THz})$, where tilde stands for the Fourier transforms, we obtain the following equation:

$$\left[ \nabla^2 + \frac{\omega_{THz}^2}{c_0^2} (n(\omega_{THz}) + i\kappa(\omega_{THz}))^2 \right] \tilde{E}_{THz}(x,y,z,\omega_{THz}) = \frac{\chi^{(2)}}{c_0^2} \int_{-\infty}^{+\infty} \frac{\partial^2}{\partial t^2} (|E_p(x,y,z,t)|^2) \cdot e^{-i\omega_{THz} t} dt \tag{11}$$

7

where $n(\omega_{THz}) + i\kappa(\omega_{THz}) = \sqrt{1 + \chi^{(1)}(\omega_{THz})}$ is the effective complex refractive index at THz frequency. The evolution of the pump envelope is described by:

$$\varepsilon(z,t) = \frac{1}{2\pi}\int_{-\infty}^{+\infty}\tilde{\varepsilon}(0,\omega_{THz})e^{i\omega_{THz}(t-\frac{z}{c_0}n_g)}d\omega_{THz} \qquad (12)$$

$\tilde{\varepsilon}(0,\omega_{THz})$ is Fourier transform of the pulse envelope at the beginning of the transmission line, $n_g$ is the group index at central pump frequency. We assume the pulse profile at the boundary z=0 has the form of:

$$\varepsilon(0,t) = \text{sech}(t/\tau) \qquad (13)$$

where $\tau \approx \frac{\text{FWHM}}{1.76}$ is the pulse duration, FWHM is the full width at half maximum determined with pulse intensity. We assume that the transmission line supports one mode, the spatial profile of which does not depend on the THz frequency so that we can express the THz electric field as

$$\tilde{E}_{THz}(x,y,z,\omega_{THz}) = \tilde{E}_{THz}^{(\text{amp})}(z,\omega_{THz}) \cdot e^{-i\frac{\omega_{THz}}{c_0}n(\omega_{THz})z} \cdot g_{THz}(x,y). \qquad (14)$$

where $g_{THz}(x,y)$ is the THz mode profile, normalized such that $\iint_{(x,y)}|g_{THz}(x,y)|^2 dxdy = S_{THz,\text{eff}}$, where $S_{THz,\text{eff}}$ is the effective mode area of THz mode. By substituting Eq. 12 into Eq. 11:

$$\left[\nabla^2 + \frac{\omega_{THz}^2}{c_0^2}(n(\omega_{THz}) + i\kappa(\omega_{THz}))^2\right]\tilde{E}_{THz}^{(\text{amp})}(z,\omega_{THz}) \cdot e^{-i\frac{\omega_{THz}}{c_0}n(\omega_{THz})z} \cdot g_{THz}(x,y) = \\ \frac{-\pi\chi^{(2)}E_0^2\omega_{THz}^3\tau^2}{2c_0^2\sinh(\pi\omega_{THz}\tau/2)}e^{-i\frac{\omega_{THz}}{c_0}n_g z} \cdot g_{opt}^2(x,y) \qquad (15)$$

To solve this equation, we apply the following simplifications: first, we neglect the derivatives with respect to x and y axis, second, we can multiply both parts of the equation by $g_{THz}^*(x,y)$ and integrate over x and y coordinates. This procedure allows us to do the analysis of 1D case, where we describe only propagation along z axis. We can introduce the overlap factor between THz and optical modes, which impacts the THz generation efficiency at a given mode:

$$\Gamma_{\text{overlap}} = \frac{\iint_{(x,y)}g_{opt}^2(x,y)g_{THz}^*(x,y)dxdy}{\iint_{(x,y)}|g_{THz}(x,y)|^2 dxdy} \qquad (16)$$

Next, we can rewrite Eq. 15 as

$$\left[\frac{\partial^2}{\partial z^2} + \frac{\omega_{THz}^2}{c_0^2}(n(\omega_{THz}) + i\kappa(\omega_{THz}))^2\right]\tilde{E}_{THz}^{(\text{amp})}(z,\omega_{THz}) \cdot e^{-i\frac{\omega_{THz}}{c_0}n(\omega_{THz})z} = \frac{-\pi\chi^{(2)}E_0^2\omega_{THz}^3\tau^2}{2c_0^2\sinh(\pi\omega_{THz}\tau/2)}e^{-i\frac{\omega_{THz}}{c_0}n_g z} \cdot \Gamma_{\text{overlap}} \qquad (17)$$

We can now simplify Eq. 17 by applying the slowly-varying amplitude approximation to the THz wave, thus neglecting the second derivative with respect to the z-coordinates. Moreover, we can assume that the THz absorption is weak $|n(\omega_{THz})| \gg |\kappa(\omega_{THz})|$, so that $(n(\omega_{THz}) + i\kappa(\omega_{THz}))^2 \approx n^2(\omega_{THz}) + 2i\kappa(\omega_{THz})n(\omega_{THz})$. We can now rearrange the term on the left side of the equation as:

$$\left[\frac{\partial^2}{\partial z^2} + \frac{\omega_{THz}^2}{c_0^2}(n(\omega_{THz}) + i\kappa(\omega_{THz}))^2\right]\tilde{E}_{THz}^{(\text{amp})} \cdot e^{-i\frac{\omega_{THz}}{c_0}n(\omega_{THz})z} \approx \\ \frac{\partial\tilde{E}_{THz}^{(\text{amp})}}{\partial z} \cdot e^{-i\frac{\omega_{THz}}{c_0}n(\omega_{THz})z} \cdot \left(-2i\frac{\omega_{THz}}{c_0}n(\omega_{THz})\right) + 2i\kappa(\omega_{THz})n(\omega_{THz})\frac{\omega_{THz}^2}{c_0^2}\tilde{E}_{THz}^{(\text{amp})} \cdot e^{-i\frac{\omega_{THz}}{c_0}n(\omega_{THz})z} \qquad (18)$$

Introducing the loss coefficient $\alpha(\omega_{THz}) = -\frac{2\omega_{THz}\kappa(\omega_{THz})}{c_0}$ and substituting this into Eq. 17 gives the equation for the evolution of the THz field:

$$\frac{\partial\tilde{E}_{THz}^{(\text{amp})}}{\partial z} = \frac{-i\pi\chi^{(2)}E_0^2\omega_{THz}^2\tau^2}{4c_0 \cdot n(\omega_{THz})\sinh(\pi\omega_{THz}\tau/2)}e^{-i\Delta kz} \cdot \Gamma_{\text{overlap}} - \frac{\alpha(\omega_{THz})}{2}\tilde{E}_{THz}^{(\text{amp})} \qquad (19)$$

where $\Delta k = \frac{\omega_{THz}}{c_0}(n_g - n(\omega_{THz}))$ is the wave vectors mismatch. Considering the boundary condition

$$\tilde{E}_{THz}^{(\text{amp})}(z=0,\omega_{THz}) = 0 \qquad (20)$$

8

We find the solution of the equation after the propagation over length L:

$$\tilde{E}_{THz}^{(amp)}(L, \omega_{THz}) = \frac{i\pi\chi^{(2)}E_0^2\omega_{THz}^2\tau^2 L\Gamma_{overlap}}{4c_0 n(\omega_{THz})\sinh(\pi\omega_{THz}\tau/2)}G_{TL}(\omega_{THz}) \quad (21)$$

where we introduce the phase-matching function $G_{TL}(\omega_{THz}) = \frac{e^{-i\Delta kL} - e^{-\alpha L/2}}{i\Delta kL - \alpha L/2}$ that quantifies the generation efficiency. In order to calculate the time trace of the THz pulse, we perform the inverse Fourier transform:

$$E_{THz}(L, t) = \int_{\omega_{THz}} \tilde{E}_{THz}^{(amp)}(L, \omega_{THz}) \cdot e^{i\omega_{THz}\left(t - \frac{Ln(\omega_{THz})}{c_0}\right)} d\omega_{THz} \quad (22)$$

And, finally, we can introduce the THz energy pulse $J_{THz}$ as:

$$J_{THz} = \int_{-\infty}^{\infty} \iint_{(x,y)} 1/2\varepsilon_0 c_0 \varepsilon_{eff} |E_{THz}(L, t)|^2 \cdot |g_{THz}(x, y)|^2 dxdydt \quad (23)$$

We note that, as discussed in [7], these expressions are valid under the following assumptions: (i) the conversion efficiency $\eta = \frac{J_{THz}}{J_{opt}} \ll \frac{\Delta\omega_p}{\omega_p}$, where $\Delta\omega_p \approx \frac{1.12}{FWHM}$ is the pump pulse bandwidth, and $\omega_p$ is pump frequency. In our case, $\frac{\Delta\omega_p}{\omega_p} \approx 5.1 \cdot 10^{-2}$, and as we show, it is much above the generation efficiencies of our device; (ii) the $\chi^{(3)}$-based effects, such as self-phase modulation, are negligible for the optical pump. Nonlinear phase, associated with the Kerr-nonlinearity, is negligible and can be estimated as $\delta\varphi \approx n_2 \cdot I_{peak} \cdot k_p l_{TL} \approx n_2 \frac{J_{opt}}{\tau S_{eff}} \frac{2\pi}{\lambda_p} l_{TL} \approx 10^{-3}$, where $I_{peak}$ is the peak intensity of the pump pulse, $k_p$ is the wave vector, and $n_2 \approx 10^{-19} m^2/W$ is the nonlinear index of $LiNbO_3$ [11].

## B Estimation of generated terahertz fields confined to the transmission line

Using the derivation of section 2 A, we investigate the impact of losses and phase matching onto the generated terahertz field amplitude at the end of a transmission line of arbitrary length. Towards this end, we first simulate the losses, overlap and refractive index for various dimensions gap widths $w_g$, shown in Supplementary Fig. 7 using CST Studio. We find that an index between $n_{TL} = 2.2 - 2.4$ can be engineered by changing the gap width, and that this influences the losses and the overlap quite significantly. First, we calculate the radiative and absorption losses, and the results are shown in Supplementary Fig. 6. To do this, we perform two simulations: one with lossy materials (both gold and $LiNbO_3$) and the second simulation with lossless materials. In the first case, the losses come from both radiation and absorption, while in the second case, losses come only from radiation. We launch the propagation over the transmission length $l_{TL} = 100$ $\mu$m and extracted the transmission T as the function of frequency. From this simulation we compute the loss coefficient as follows:

$$\alpha(f_{THz}) = -\frac{1}{l_{TL}}\ln(T(f_{THz})) \quad (24)$$

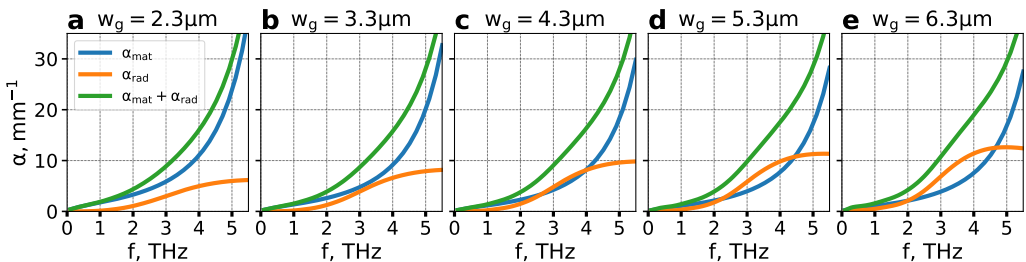

**Supplementary Fig. 6: Radiative and absorption losses of the tranmission line**

The difference between the total loss coefficient $\alpha = \alpha_{\text{mat}} + \alpha_{\text{rad}}$ and radiative loss coefficient $\alpha_{\text{rad}}$ gives the absorption coefficient $\alpha_{\text{mat}}$, which is plotted in Fig. 6. One may observe that the radiative losses are increasing for larger gap width, while the absorption losses are decreasing, since the THz mode spatial overlap with the absorbing materials (LiNbO$_3$) is decreasing.

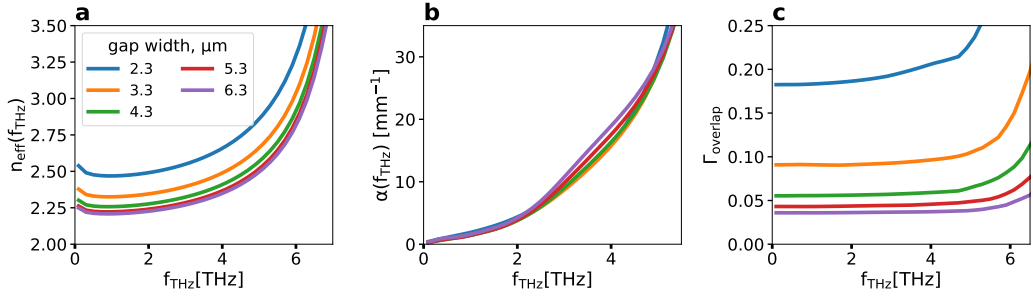

**Supplementary Fig. 7: Properties of the transmission lines for various gap widths as a function of frequency** **a** Effective index of the transmission line n$_{\text{TL}}$. **b** Absorption losses inside the transmission line. **c** Overlap integral as defined in section A.

Accounting for all of these parameters, we compute the phase matching function G($\omega_{\text{THz}}$) while remarking that, generally, the terahertz field amplitude is dependent as

$$\tilde{\mathrm{E}}_{\text{THz}}^{(\text{amp})}(l_{\text{TL}}, \omega_{\text{THz}}) \sim \chi^{(2)} \mathrm{E}_0^2 \omega_{\text{THz}}^2 l_{\text{TL}} \Gamma_{\text{overlap}} \mathrm{G}_{\text{TL}}(\omega_{\text{THz}}) \tag{25}$$

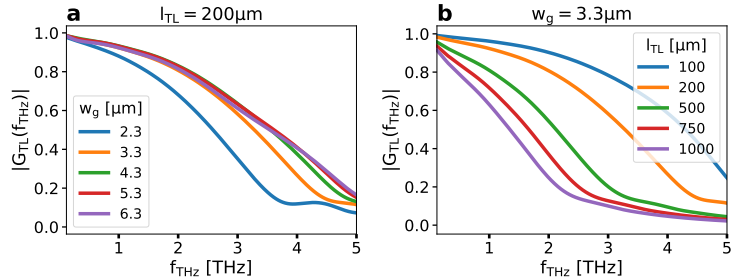

**Supplementary Fig. 8: Phase matching function. a** For a fixed transmission line length of l$_{\text{TL}} = 200$ $\mu$m and with varying transmission line width. **b** For a varying transmission line length and with fixed transmission line width of w$_g = 3.3$ $\mu$m, as in our experiments.

since n$_{\text{TL}}(\omega_{\text{THz}})$ is rather flat with frequency. We note that a long transmission line only practical if the total product is larger than the one for a short interaction length. To get an intuition for this, we plot in Supplementary Fig. 8 the phase matching function for varying gap widths and varying transmission line lengths. In general, we find that a transmission line of l$_{\text{TL}} = 200$ $\mu$m can provide proper phase matching for frequencies surpassing 3 THz for widths above w$_g = 3$ $\mu$m and that a transmission line of length l$_{\text{TL}} = 1$ mm is adequate at frequencies below 1 THz.

The terahertz electric field at the end of a transmission line of given length L = l$_{\text{TL}}$ is evaluated using equations 21-22 and the results are presented in Supplementary Fig. 9. For this calculations, we choose the experimental parameters of the optical pump pulse: for a pump pulse energy of J$_{\text{opt}} = 1$ pJ and pulse length of $\tau \approx \frac{\text{FWHM}}{1.76} \approx 60$ fs. First, using equation 21 we analyzed the generated spectra at the end of the transmission line which are shown in Fig. 9a and Fig. 9b. As expected, for longer propagation lengths, the spectrum becomes narrower due to the both

10

absorption and phase-matching conditions. Also, varying the gap between two metallic strips of the transmission line, we show, that the biggest amplitude and the largest bandwidth is achieved for the smallest gap. This is due to the better overlap between optical and terahertz modes and phase-matching conditions. Next, we evaluated the waveforms, using equation 22 and results are shown in subplots Fig. 9c and Fig. 9d. We find exemplary for a transmission line length of $l_{TL} = 200$ $\mu$m that electric fields on the order of $E_{THz} = 1$ kV/cm are generated at the end of the transmission line for different gap widths and transmission line lengths, as it is summarized in Fig. 9e.

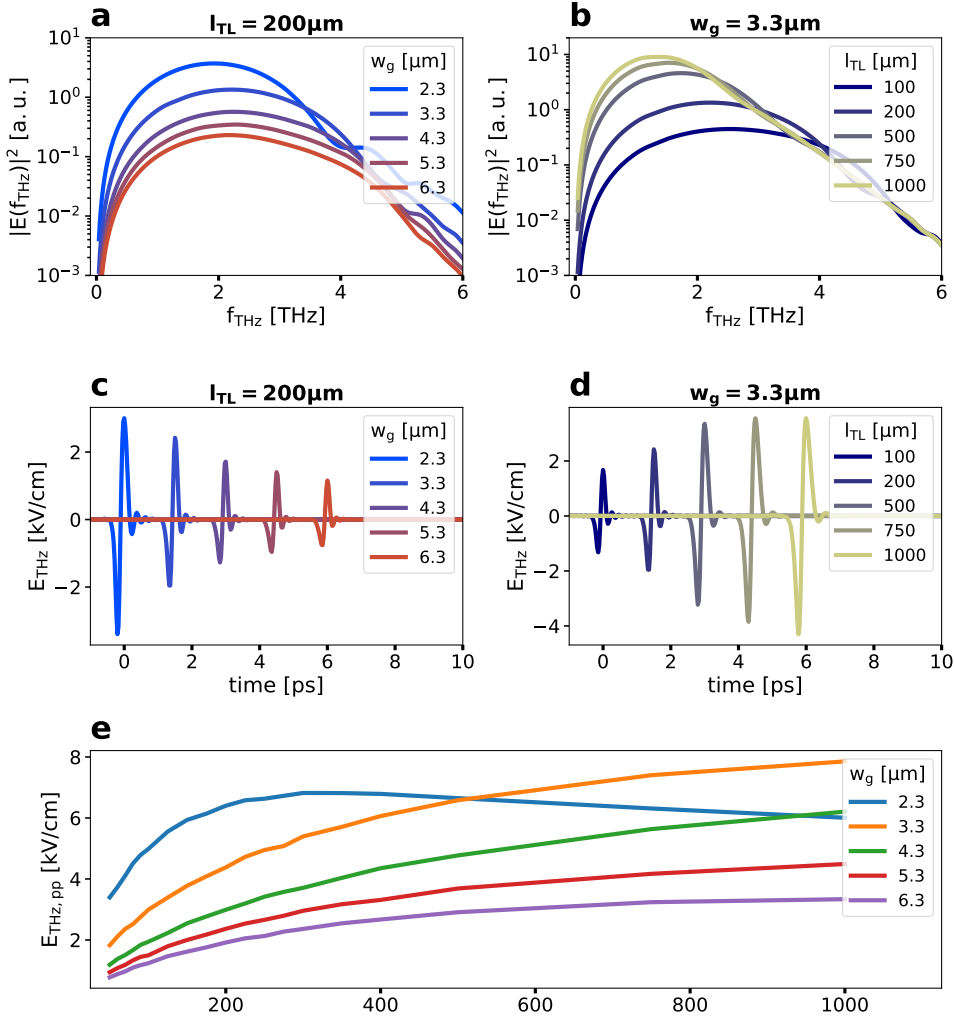

**Supplementary Fig. 9: Terahertz electric field at the end of the transmission line as estimated from the analytical model of section A**. Calculated spectra **a** and time transients **c** for various gap widths $w_g$ and fixed transmission line length $l_{TL}$; Calculated spectra **b** and time transients **d** for fixed $w_g$ and various $l_{TL}$; **e** Peak-to-peak values of the generated THz pulses as a function of length and different waveguide widths.

We estimated the efficiency of the THz pulse generation $\frac{J_{THz}}{J_{opt}^2}$ to be $\sim 10^{-4}$/pJ, as shown in Fig. 10. We note, that this value does not depend on pump pulse energy, and it gives a conversion efficiency of $\eta = \frac{J_{THz}}{J_{opt}} \sim 10^{-4}$ for pump pulse energy $J_{opt} = 1$ pJ.

## C  Modelling transmission line cavity

In this section, we model the cavity formed by the transmission line discussed in Fig. 4 in the main text. Towards this end, we rely on the derivation of the generated terahertz field of section A, and on the reflection $S_{11}$ and loss $\alpha$

11

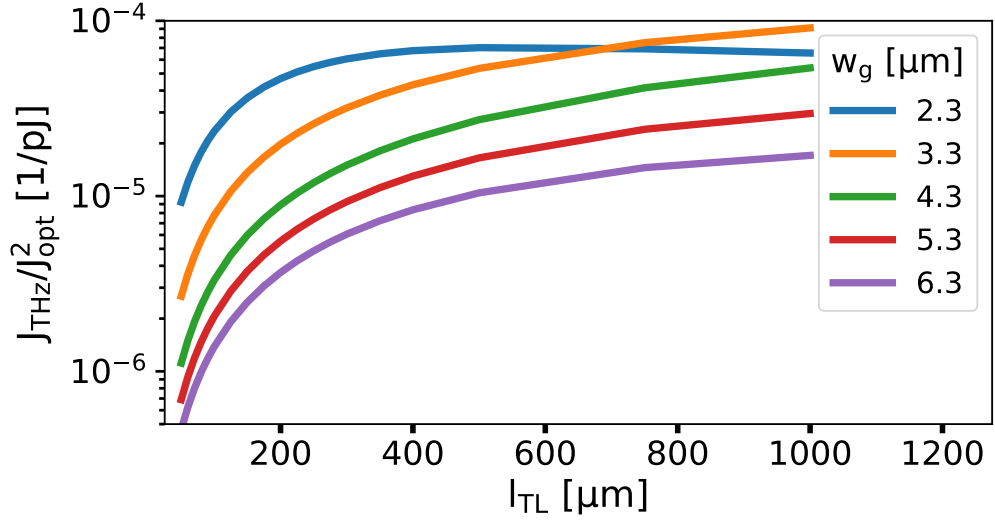

**Supplementary Fig. 10: Conversion efficiency of the generation of THz pulses estimated from the analytical model of section A**

parameters of Fig. 7 we have determined for the transmission lines.

The total outcoupled field $E_{out}$ is a sum over the various contributions of the terahertz pulses travelling back and forth inside the transmission line, as sketched in Fig. 11a:

$$E_{out} = E_{out}^{(1)} + E_{out}^{(2)} + E_{out}^{(3)} + ... \tag{26}$$

where we have:

$$E_{out}^{(1)} = t_a \tilde{E}_{THz}^{(amp)}\left(\frac{l_{TL}}{2}, \omega_{THz}\right) e^{-i\frac{\omega_{THz}}{c} n(\omega_{THz}) \frac{l_{TL}}{2}} \tag{27}$$

$$E_{out}^{(2)} = t_a \cdot \left( S_{11} \cdot \tilde{E}_{THz}^{(amp)}\left(\frac{l_{TL}}{2}, \omega_{THz}\right) e^{-i\frac{\omega_{THz}}{c} n(\omega_{THz}) l_{TL}} e^{-\frac{\alpha(\omega_{THz})}{2} \cdot \frac{l_{TL}}{2}} + r_a S_{11} E_{out}^{(1)} e^{-\frac{\alpha(\omega_{THz}) l_{TL}}{2}} e^{-i\frac{\omega_{THz}}{c} n(\omega_{THz}) l_{TL}} \right) \tag{28}$$

$$E_{out}^{(3)} = t_a \cdot S_{11} E_{out}^{(2)} e^{-\frac{\alpha(\omega_{THz}) l_{TL}}{2}} e^{-i\frac{\omega_{THz}}{c} n(\omega_{THz}) l_{TL}} \tag{29}$$

$$... \tag{30}$$

$$E_{out}^{(i+1)} = t_a \cdot S_{11} E_{out}^{(i)} e^{-\frac{\alpha(\omega_{THz}) l_{TL}}{2}} e^{-i\frac{\omega_{THz}}{c} n(\omega_{THz}) l_{TL}} \tag{31}$$

By assuming the extraction of the antenna to be $t_a = 0.3$ and $t_{TL} = 0.9$, we find a very good agreement between experimental data and modelling, as illustrated in Supplementary Fig. 11b-c. The time-domain signatures featuring several reflections, and the well-resolved frequency-domain resonances, confirm the emergence of cavity modes through repeated reflections at the open end termination of the transmission line. The constructive interference of these fields locally enhances the terahertz electric field at the resonance condition compared to a single-pass field.

### D  Dispersion of THz transmission line

Since our design provides a relatively large coherence length, group velocity dispersion (GVD) of the THz mode is calculated. The group index of THz is calculated from the effective refractive index values in the CST simulations. Experimentally, the value is estimated from the free spectral range (FSR) of the cavity modes in Fig. 4 in the main text using the formula

$$\Delta\nu_{FSR} = \frac{c_0}{2n_g \cdot l_{TL}} \tag{32}$$

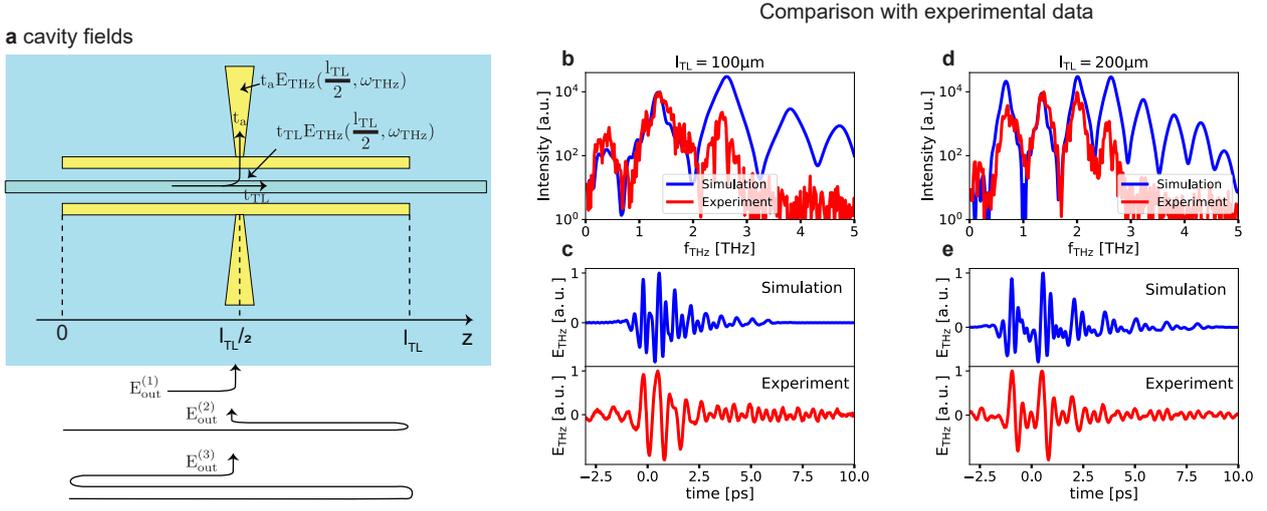

**Supplementary Fig. 11: Modelling of the cavity properties. a** Sketch of the cavity formed by a transmission line with a centrally patterned extracting terahertz antenna. The total outcoupled terahertz electric field is the sum of the outcoupled fields after various round-trips inside the cavity $E_{\text{out}}$. Comparison between experimental data in main text 4e for the transmission line length of **b-c**: 100 $\mu m$; **d-e**: 200 $\mu m$. Results show good agreement in the number of pulses, as well as in the location of the Fabry-Perot resonances and depth of the dips, confirming the existence of cavity modes.

and it lies in the range 2.1-2.7 which agrees with our simulations. The GVD at a central THz wavelength $\lambda_{\text{THz},0}$ is determined from the simulated refractive index.

$$\text{GVD} = \frac{\lambda_{\text{THz},0}^3}{2\pi c^2} \left( \frac{\partial^2 n}{\partial \lambda_{\text{THz}}^2} \right)_{\lambda_{\text{THz}} = \lambda_{\text{THz},0}} \tag{33}$$

The GVD of our device is compared to the bulk Lithium Niobate (Supplementary Fig. 12 b). Our structure provides

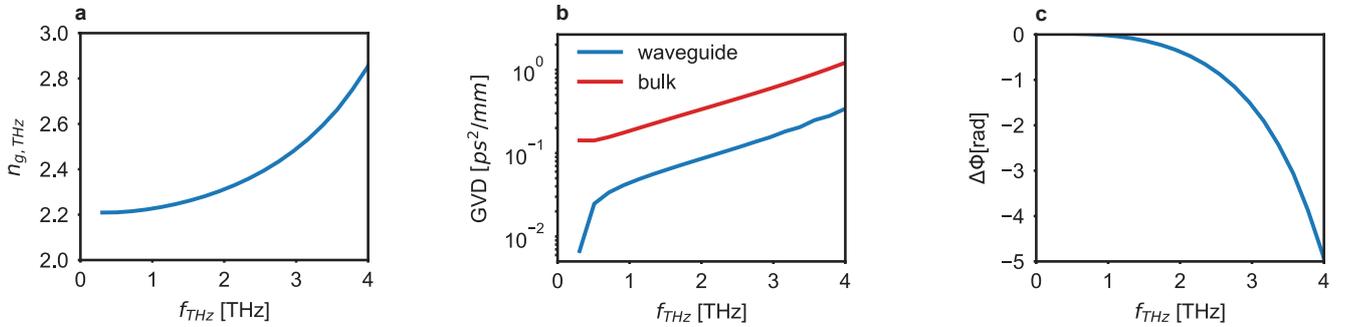

**Supplementary Fig. 12: a** Group index of THz mode. **b** Group velocity dispersion (GVD) of THz radiation in the case of bulk Lithium Niobate and our transmission line mode. **c** Change in phase that originates from propagation in a 120 $\mu m$ THz transmission line.

around factor 4 decrease in GVD compared to bulk Lithium Niobate. Next, we analyze the effect of the propagation distance $l_{\text{TL}}$ and dispersion on the carrier envelope offset $\phi_{\text{ceo}}$ of the THz pulse. This can be calculated using the formula [12]

$$\phi_{\text{ceo}} = l_{\text{TL}} 2\pi \left( \frac{\partial n}{\partial \lambda_{\text{THz}}} \right)_{\lambda_{\text{THz}} = \lambda_{\text{THz},0}} \tag{34}$$

We use 120 $\mu$m to predict the effect of the transmission line on the THz pulse. As depicted in Supplementary Fig. 12 (b), the effect becomes relevant at frequencies above 2 THz. At 3.5 THz, a complete phase flip ($\Delta\phi = 180$ degrees) occurs.

13

### E  Time-delayed terahertz dipole sources

To visualize the generation of terahertz radiation from the device in Fig 2 in the main text ($l_{ant} = 200$ $\mu$m and $l_{TL} = 120\mu$m) in time during propagation of the pump inside the transmission line and during the emission into farfield, 39 subsequent dipole sources are created along the Lithium Niobate waveguide (Supplementary Fig.13 a and b). The excitation of each source $m$ is delayed by $\Delta t_m = m \cdot n_g \cdot d/c_0$, where $d = 10$ $\mu$m is the distance between each two subsequent dipole sources. The results of the time domain simulation are shown in the top and side view in 13 c and d with 500 fs time steps. To represent the terahertz generation and guiding, the sources are first excited at position -180 $\mu m$ where the pump has not yet reached the transmission line. As anticipated, up to the time step 2.5 ps, the generated terahertz is not guided and emission into the substrate is efficient at the Cherenkov angle $\theta_c = \arcsin \frac{n_g}{n_{THz}} \approx 42°$ (highlighted at time 2.5 in d). Once the dipole sources inside the transmission line are excited, the generated terahertz radiation is guided towards the antenna (time 3.5 ps in c). While the real substrate thickness is 500 $\mu m$, 100 $\mu m$ is chosen for the simulation to reduce required memory and computation. The reflection at the substrate-air interface is shown in d at time 5 ps where the diverging terahertz radiation will be collected by the parabolic mirror (shown in the optical setup in Supplementary Fig. 3).

### F  Farfield analysis

The divergence of the THz field into air as shown in Supplementary Fig. 13 d after time 4.5 ps necessitates analysis of the antenna's radiation pattern at different frequencies to estimate the collection efficiency (Supplementary Fig. 14). Due to the broadband nature of our THz device, the following analytical expression is used to calculate the normalized radiation intensity in the farfield as a function of emission angle $\theta$ [13]:

$$U(\theta, f_{THz}) = \left[ \frac{\cos(\beta_{THz} l_{ant} \cos\theta) - \cos(\beta_{THz} l_{ant})}{\sin\theta} \right]^2 \tag{35}$$

The total power for each frequency can be then calculated by integrating over the all angles

$$P_{total}(f_{THz}) = \int_0^\pi \int_0^{2\pi} U(\theta, f_{THz}) \sin\theta d\phi d\theta \tag{36}$$

Next, the collection angle $\alpha_{coll}$ is determined from the specifications of the used parabolic mirror:

$$\alpha_{coll} = \arctan(a/RFL) \tag{37}$$

where a is the radius and RFL is the reflective focal length of the parabolic mirror. In our case the angle is $\alpha_{coll} = 26.6$ degrees. The corresponding angle in the substrate $\beta$ is $\beta = \arcsin\left(\frac{\sin \alpha_{coll}}{n_{Si}}\right) = 7.55$ degrees. In Supplementary Fig. 15 a, the farfield radiation pattern is plotted using equation 35 for different frequencies and for our antenna $l_{ant} = 200$ $\mu$m. Next, the intensity is integrated over the collection angle

$$P_{collected}(f_{THz}) = \int_{\pi/2-\beta}^{\pi/2+\beta} \int_{-\beta}^{\beta} U(\theta, f_{THz}) \sin\theta d\phi d\theta \tag{38}$$

The results of equations 38 and 36 are shown in Fig. 15 b for three different antenna lengths. While the total radiated power is similar for different antenna lengths, collection of the radiation strongly depends on the radiation pattern. It is evident that our antenna provides a flat response at frequencies above 1 THz compared to shorter antennas despite the overall decrease in collected power after reaching the resonance. The collection efficiency of the emitted electric field can be written as

$$\eta_{coll} = \sqrt{\frac{P_{collected}}{P_{total}}} \tag{39}$$

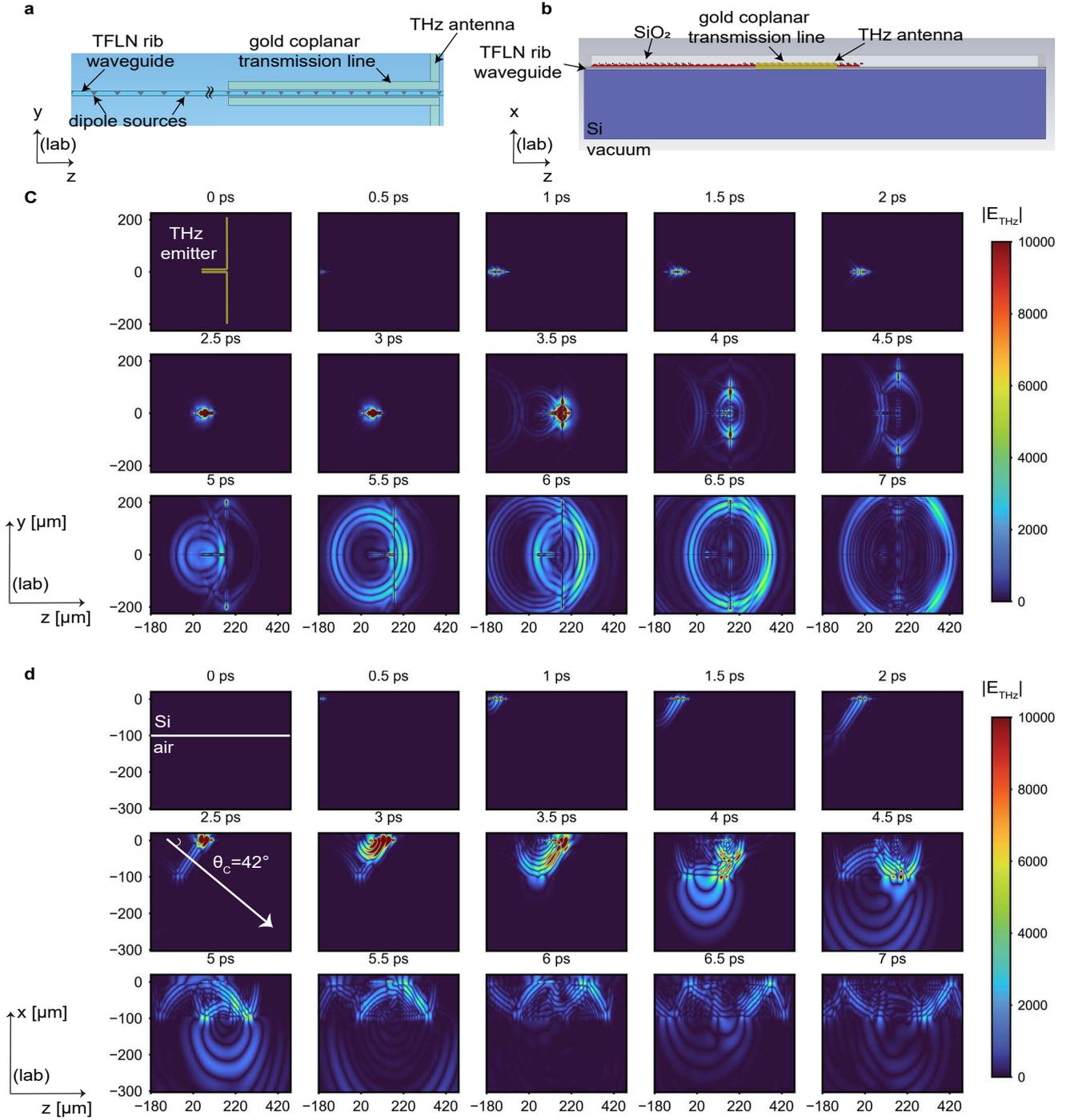

**Supplementary Fig. 13: Time-domain CST simulation of delayed THz dipole sources for the dipole antenna in Fig. 2 in the main text.** Simulated structure in CST viewed from the top (**a**) and from the side (**b**). **c** top view of the generated $|E_{THz}|$ from the dipole sources with 500 fs time steps. **d** side view of the THz field showing: Cherenkov radiation up to 2.5 ps, confinement of the THz in the transmission line, emission from the antenna into the substrate, reflection at the substrate-air interface and emission into the farfield.

We find that $\eta_{\text{coll}} \sim 0.05$. To show that the THz beam is illuminating the parabolic mirror completely, a knife edge measurement is conducted with 1 mm steps on a device with antenna length $l_{\text{ant}} = 200$ μm. The blocking surface is placed in the collimating path after the first parabolic mirror. As clearly seen in Supplementary Fig. 16, a change in amplitude is evident along the complete range of the parabolic mirror.

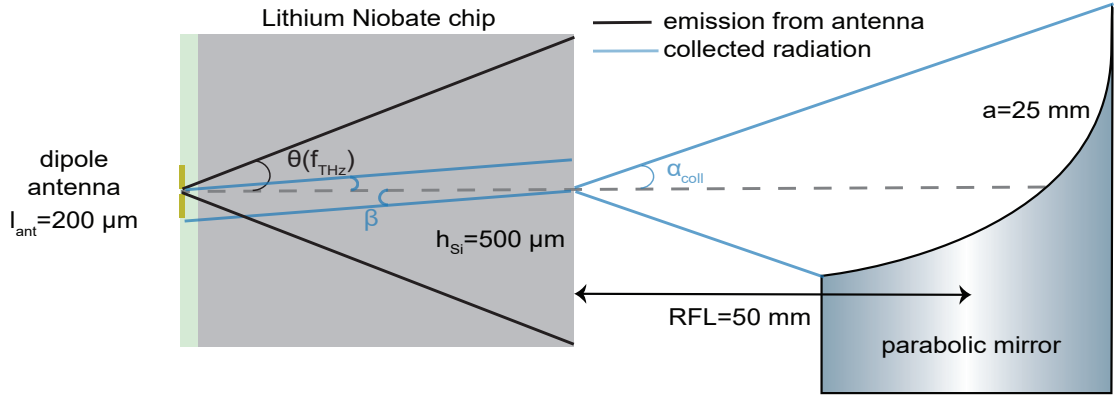

**Supplementary Fig. 14: Collection of the emitted THz radiation by the parabolic mirror**. The dipole antenna emits into the Silicon substrate at a frequency-dependent angle $\theta(f_{\text{THz}})$. Due to diffraction from Silicon to air and due to the limited area of the parabolic mirror, only radiations at angle $\beta$ can be collected. $\alpha_{\text{coll}}$ is the the collection angle of the parabolic mirror, RFL: reflective focal length.

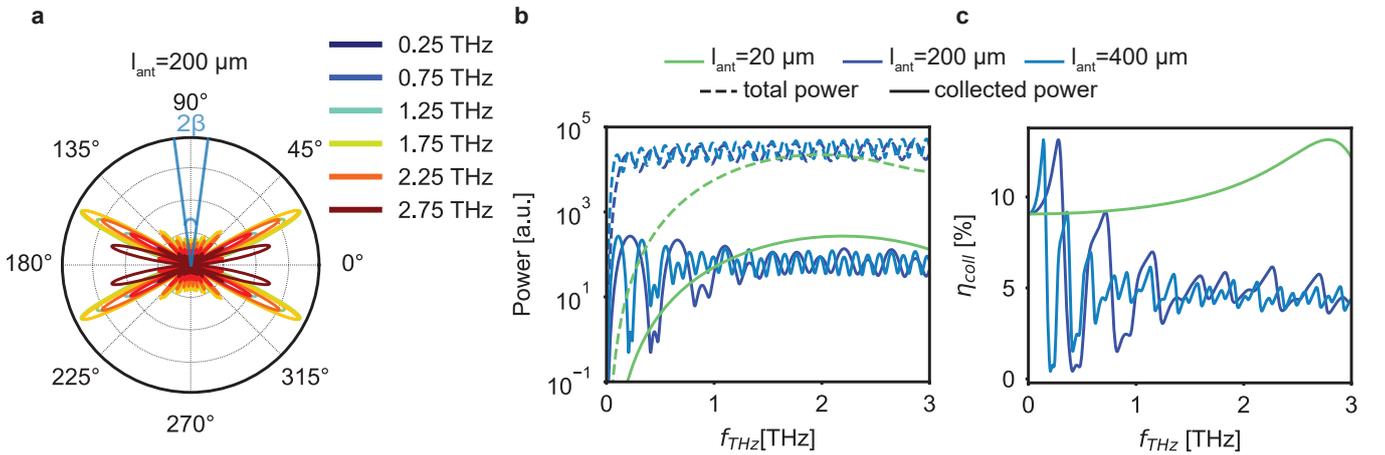

**Supplementary Fig. 15: Farfield radiation patterns of our dipole antenna and collected power from the first parabolic mirror.** **a** radiation pattern of a dipole antenna for different THz frequencies. The blue line is the collection angle inside the substrate which is constrained by the refractive index of the substrate. **b** the total and collected power values using equations 38 and 36 for different antenna length.

16

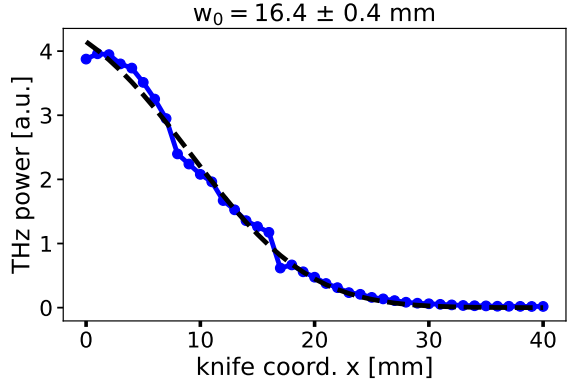

**Supplementary Fig. 16:** knife edge measurement of the collimated THz beam emitted from the dipole antenna in Fig. 2. Dashed line is the fitting of the power to the erf function.

## 3 Further supporting measurements of chip-scale emitters

### A Estimation of collection and detection efficiencies

The emitted THz electric field from the antenna will encounter the following amplitude reflection coefficients:

$$r_{\text{total}} = r_{\text{Si}} \cdot r_{\text{crystal}} \tag{40}$$

$r_{\text{Si}}$ is the reflection of THz at the Silicon substrate air interface. Neglecting dispersion and assuming perpendicular incidence, its absolute value is 0.54. For the detection crystals, $r_{\text{crystal}}$ amounts to 0.54 for ZnTe. The total reflection is in the orders of 0.29 and 0.28 for the crystals accordingly. One further contribution for the undetected THz is the mismatch between the mode area inside the transmission line and the area of the focused THz beam inside the crystal. Our simulations in Sec 2 A give a mode area of $S_{\text{eff,THz}} = 6\ \mu m^2$ for the measured devices with gap of 3.3 $\mu$m. The waist of the beam at the facet of the detection crystal can be calculated from

$$2w_{\text{crystal}} = \frac{4 \cdot c}{f_{\text{THz}} \cdot \pi} \cdot \frac{\text{RFL}}{2w_0} \tag{41}$$

and yields with RFL = 50 mm and $2w_0 = 32$ mm a beam waist at the crystal facet of $2w_{\text{crystal}} = 596\ \mu$m, corresponding to an area of $A_2 = \pi/2 \cdot w_{\text{crystal}}^2 = 139 \cdot 10^3 \mu m^2$ at 1 THz. The Area mismatch is given as

$$A = \frac{S_{\text{eff,THz}}}{A_2} \tag{42}$$

and at 1 THz, it amounts to $4.3 \cdot 10^{-5}$. As derived in A and shown in Supplementary Fig. 9, we estimate a peak THz amplitude in the order of $600 \cdot 10^3$ V/m at the end of the transmission.

The detected THz field can be written as

$$E_{\text{det}} = E_{\text{THz}} \cdot \eta_{\text{coll}} \cdot r_{\text{total}} \cdot \sqrt{A} \tag{43}$$

where $E_{\text{THz}}$ is the electric field in time domain derived in equation 21. Collection of the THz radiation by the optics and the antenna's farfield characteristics are included in $\eta_{\text{coll}}$ equation 39. The reflection $r_{\text{total}}$ is described in equation 40 and the area mismatch in equation 42. We estimate the terahertz electric field detected in the farfield to be approximately $\eta_{coll} \cdot r_{total} \cdot \sqrt{A} = 0.05 \cdot 0.28 \cdot \sqrt{4.3 \cdot 10^{-5}} \approx 10^{-4}$ lower than the one at the end of the transmission line. Given that for pump pulse energies of 1 pJ we estimate field amplitudes inside the transmission line of $100 \cdot 10^2$ V/m, corresponding to $100 \cdot 10^4$ V/m for 100 pJ pulse energies, this is consistent with our measurements

17

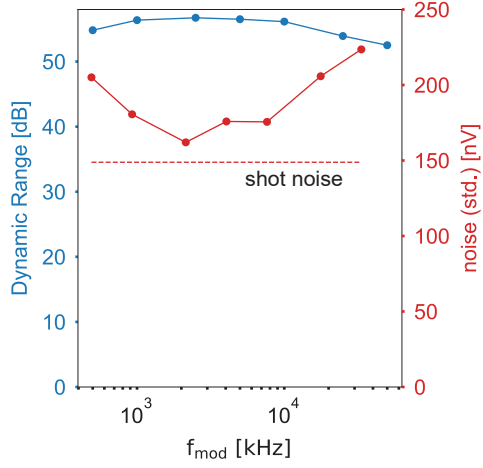

**Supplementary Fig. 17:** Noise and dynamic range with modulation frequency. Dashed line describes the shot noise contributions

with field amplitudes on the order of 100 V/m for pump pulse energies of 100 pJ. We compute the experimental efficiency of terahertz generation using the formula $\eta = \frac{J_{opt}}{J_{THz}}$, with $J_{opt} = 77$ pJ and $J_{THz} = \frac{1}{2}c_0 n\epsilon_0 E_{THz}^2 A_2 \tau_{THz}$ where $E_{THz} = 43$V/m, $\tau_{THz} = 800$ fs is the terahertz pulse length and find $J_{THz} = 3.5 \cdot 10^{-19}$ J. Hence, the as-measured generation efficiency is about $\eta_{as-measured} = 4.5 \cdot 10^{-9}$. We can estimate the actual generation efficiency inside the transmission line $\eta_{TL,estimated} = \frac{\eta_{as-measured}}{\eta_{coll}^2 \cdot r_{total}^2} = 2.3 \cdot 10^{-5}$.

### B  Shot-noise limited detection using electro-optic sampling

In order to show that the detection scheme approaches the shot-noise limit at high modulation frequency, the fundamental limit is calculated from the experimental conditions using the equation

$$\delta V_{shot} = \sqrt{4eV_0 \frac{G}{\Re} \Delta f} \qquad (44)$$

where $V_0$ is the voltage on the photodetector, e is the electron's charge, G is the conversion gain in units of V/W, $\Re$ is the responsivity in units of A/W, and $\Delta f$ is the detection bandwidth, it can be calculated from the chosen time constant for lock-in detection and the filter-order. Note, that additional factor of 2 comes from the assumption, that the shot noise from both photodetectors is not correlated. The noise calculations using the formula above and the parameters are summarized in the table 3.

| G | $\Re$, | $V_0$ | $\Delta f$ | $\delta V_{shot}$ |
|---|---|---|---|---|
| $16 \cdot 10^3$ V/W | 0.5 A/W | 8 V | 0.135 Hz | 148 nV |

**Supplementary Table 3:** Parameters used to estimate the theoretical shot noise amplitude and comparison with experimental result

We compare this calculated shot noise limit to the experimental one. Towards this end we modulate the emitter at various frequencies between 75 kHz and 50 MHz as shown in Fig. 17. Thanks to overcoming the flicker noise at MHz frequencies, we approach the shot noise limit with weak dependency on the frequency. The dynamic range is determined from Fourier spectrum of the measured THz signals. First, the maximum power spectral density is determined $E_{max}(f_{THz}) = \max(|E(f_{THz})|)$. Next, the noise power is determined in the frequency domain from taking the average of the power spectral density before the arrival of the THz signal $t_0$: $N = <|E(f_{THz})|^2>; t < t_0$. Finally,

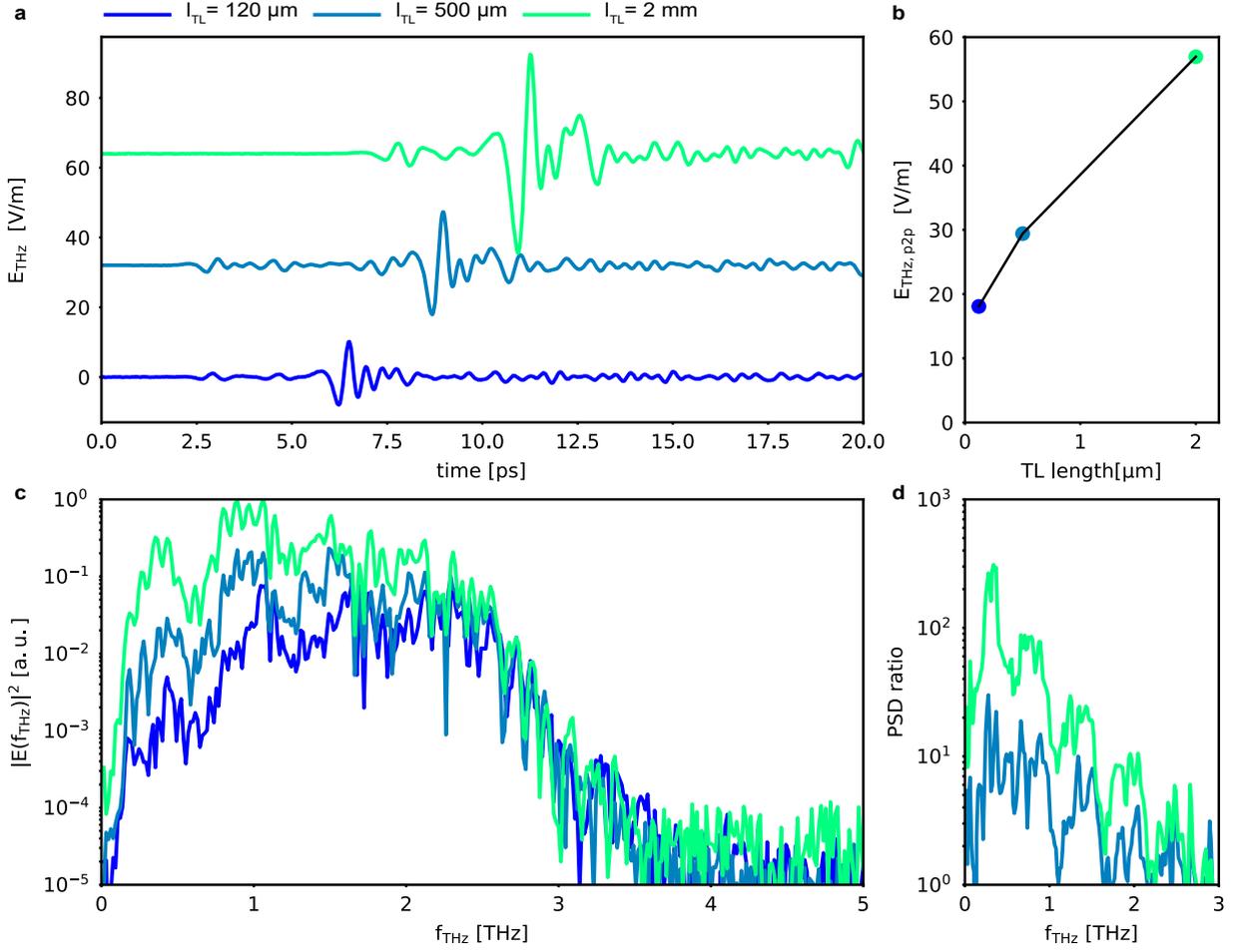

Supplementary Fig. 18: **Dependence of THz emission on the length of the transmission line $l_{TL}$ a** detected emission from three devices with same antenna length $l_{ant} = 200$ μm and varying transmission line length $l_{TL} = 120$ μm, 500 μm, 2 mm. For electro-optic sampling, ZnTe crystal is used. **b** extracted peak to peak values of the time domain traces. **c** spectra of the THz signal. **d** ratio of the power spectral densities compared to the device with $l_{TL} = 120$ μm

the dynmic range is calculated using the formula $DR = E_{max}^2/N$. The results are shown in Fig. 17. We obtain dynamic range values above 50 dB upto 50 MHz modulation frequencies.

### C    Effect of transmission line length

In the following, we experimentally demonstrate the effect of the transmission line length $l_{TL}$ on the THz generation using the setup explained in Sec. 1 C with ZnTe crystal. We investigate the emission from three devices with same antenna length $l_{ant} = 200$ μm and different transmission lines $l_{TL} = 120$ μm, 500 μm, 2 mm. The results of the measurements and their peak-to-peak values are depicted in Fig. 18 a,b. In contrast to our analytical simulations in Fig. 9, the THz peak amplitude continue to increase with longer transmission lines. The spectra of the THz signals are shown in Fig. 18 c. The enhancement in emission depends strongly on frequency. We quantify this trend by calculating the ratio of the power spectral densities compared to the device with $l_{TL} = 120$ μm (Fig. 18 d). In qualitative agreement with our simulation results in Fig 9 b, lower frequencies are greatly enhanced, due to phase matching and low absorption. Frequencies above 2.2 THz are limited by phase matching and loss, limiting THz generation with longer transmission lines.



\end{document}